\documentclass[11pt,a4paper]{article}
\pdfoutput=1
\usepackage{jheppub}
\usepackage{multirow}
\usepackage{graphicx}
\usepackage{subfigure}
\usepackage{comment}

\allowdisplaybreaks

\def \bea {\begin{eqnarray}}
\def \eea {\end{eqnarray}}

\graphicspath{{figures/}}

\begin{document}

\thispagestyle{empty}
\vspace*{2.5cm}
\vspace{0.5cm}

\title{
	Prospects of gravitational waves in the minimal left-right symmetric model
}

\author[a]{Mingqiu Li,}
\author[a,b]{Qi-Shu Yan,}
\author[c,d]{Yongchao Zhang,}
\author[b]{Zhijie Zhao}

\affiliation[a]{School of Physics Sciences, University of Chinese Academy of Sciences, Beijing 100049, China}

\affiliation[b]{Center for Future High Energy Physics, Institute of High Energy Physics, Chinese Academy of Sciences, Beijing 100049, China}

\affiliation[c]{School of Physics, Southeast University, Nanjing 211189, China}

\affiliation[d]{Department of Physics and McDonnell Center for the Space Sciences,  Washington University, \\
St.\ Louis, MO 63130, USA}

\emailAdd{limingqiu17@mails.ucas.ac.cn}
\emailAdd{yanqishu@ucas.ac.cn}
\emailAdd{zhangyongchao@seu.edu.cn}
\emailAdd{zhaozhijie@ihep.ac.cn}



\abstract{
The  left-right symmetric model (LRSM) is a well-motivated framework to restore parity and implement seesaw mechanisms for the tiny neutrino masses at or above the TeV-scale, and has a very rich phenomenology at both the high-energy and high-precision frontiers. In this paper we examine the phase transition and resultant gravitational waves (GWs) in the minimal version of LRSM. Taking into account all the theoretical and experimental constraints on LRSM, we identify the parameter regions with strong first-order phase transition and detectable GWs in the future experiments. It turns out in a sizeable region of the parameter space,  GWs can be generated in the phase transition with the strength of $10^{-17}$ to $10^{-12}$ at the frequency of 0.1 to 10 Hz, which can be detected by BBO and DECIGO. Furthermore, GWs in the LRSM favor a relatively light $SU(2)_R$-breaking scalar $H_3^0$, which is largely complementary to the direct searches of a long-lived neutral scalar at the high-energy colliders. It is found that the other heavy scalars and the right-handed neutrinos in the LRSM also play an important part for GW signal production in the phase transition.
}


\maketitle

\section{Introduction}
\label{intro}

The discovery of a Higgs boson at the Large Hadron Collider (LHC) heralds the completion of the standard model (SM)~\cite{Aad:2012tfa,Chatrchyan:2012xdj} and a great hope for the discovery of new physics. Obviously, the completion of the SM naturally leads to the quest of microscopic structure to its next chapter, which will be further searched by the LHC \cite{Morrissey:2009tf}. In the long list of questions which might be the key to the next chapter, a few are interesting and crucial. For example, what is the dynamics for the electroweak (EW) symmetry breaking, what is the origin of mass of neutrinos \cite{Mohapatra:2006gs}, how are the parity and CP symmetries broken, and what is nature of dark matter and dark energy \cite{Sahni:2004ai}, etc. To answer these questions has been motivating various new physics models beyond the SM (BSM) at the TeV scale.

In the history of early universe, from the Planck time to today, phase transitions might have occurred when the symmetries at different energy scales are broken. For example, the symmetry breaking of grand unified theory (GUT) and supersymmetry (SUSY) breaking can induce the corresponding phase transitions at the GUT scale and SUSY breaking scale.
For new physics beyond the SM, new dynamics and a larger symmetry are usually introduced at the TeV region or a higher-energy scale.  Such new physics models are of special interests, as they might accommodate baryogenesis and thus explain the matter-antimatter asymmetry observed in the universe \cite{Cohen:1993nk,Rubakov:1996vz,Trodden:1998ym,Morrissey:2012db}. Furthermore, some of the new physics models are within the reach of the LHC and the future high-energy colliders, such as the International Liear Collider (ILC)~\cite{Baer:2013cma}, Circular Electron-Positron Collider (CEPC)~\cite{CEPC-SPPCStudyGroup:2015csa}, Future Circular collider (FCC-hh)~\cite{FCC-hh} and Super Proton-Proton Collider (SPPC)~\cite{Tang:2015qga}.

First-order phase transition (FOPT) can fulfil one of the Sakharov's conditions for successful baryogenesis~\cite{Sakharov:1967dj}. One of byproduct of strong FOPT is a sizeable production of gravitational waves (GWs). The production of GWs include three physics processes~\cite{Cai:2017cbj}: bubble collision~\cite{Kosowsky:1991ua, Kosowsky:1992vn, Huber:2008hg, Kosowsky:1992rz, Kamionkowski:1993fg, Caprini:2007xq}, acoustic wave production~\cite{Hindmarsh:2013xza, Giblin:2013kea, Giblin:2014qia, Hindmarsh:2015qta}, and chaotic magnetohydrodynamic (MHD) turbulence~\cite{Caprini:2006jb, Kahniashvili:2008pf, Kahniashvili:2008pe, Kahniashvili:2009mf, Caprini:2009yp}. In the non-runaway scenario, the GWs of acoustic wave production is the dominant one. The strong FOPTs caused by new physics can produce a significant magnitude of GWs \cite{Grojean:2006bp,Ellis:2019oqb}, which can be probed by the proposed GW experiments TianQin~\cite{Luo:2015ght}, Taiji~\cite{Guo:2018npi}, LISA~\cite{Audley:2017drz, Cornish:2018dyw}, ALIA~\cite{Gong:2014mca}, MAGIS~\cite{Coleman:2018ozp}, DECIGO~\cite{Musha:2017usi}, BBO~\cite{Corbin:2005ny}, Cosmic Explorer (CE)~\cite{Evans:2016mbw}, Einstein Telescope (ET)~\cite{Punturo:2010zz}, aLIGO~\cite{LIGOScientific:2019vkc} and aLIGO+~\cite{aLIGO+}.


Since the successful detection of GWs produced by the merging of two massive objects~\cite{Abbott:2016blz, TheLIGOScientific:2017qsa}, direct GW detection has been established as a novel method to probe the early universe. Furthermore, the direct detection of thermal GWs becomes accessible to probe phase transitions of the early universe in the multi-messager era \cite{Meszaros:2019xej}. Compared to the chirp-like GW signals from the merge of massive objects which have clear sources and can most exist in a short period, the thermal GW signal is continuous, isotropic, and lasting for a very long time. Generally speaking, its peak frequencies are intimately related to the dynamics of phase transition \cite{Dev:2016feu,Weir:2017wfa}. This opens up an active and interesting study to explore phase transitions of a new physics beyond the SM at the TeV-scale and the corresponding signals at colliders and GW detectors. For example, such a study has been conducted in the effective field theory method \cite{Huang:2016odd,Huang:2016cjm}. The condition of the strong FOPT in the new physics beyond the SM can be more easily realized when the Higgs sector includes more scalars \cite{Ivanov:2017dad}. For example, there are works on a singlet extension of the SM~\cite{Vaskonen:2016yiu,Beniwal:2017eik,Alves:2018jsw,Chen:2019ebq} or more than one singlet extension~\cite{Kakizaki:2015wua, Hashino:2016rvx,Hashino:2016xoj,Kang:2017mkl}, two-Higgs-doublet models (2HDMs)~\cite{Cline:1996mga,Basler:2016obg,Dorsch:2016nrg,Huang:2017rzf} or other doublet extensions~\cite{Wang:2019pet,Paul:2020wbz}, models with triplet extension~\cite{Chala:2018opy}, SUSY models~\cite{Apreda:2001us,Huber:2015znp,Huber:2007vva,Demidov:2017lzf}, composite models~\cite{Chala:2016ykx,Bruggisser:2018mrt,Bian:2019kmg} and walking technicolor models~\cite{Jarvinen:2009mh, Chen:2017cyc,Miura:2018dsy}, twin Higgs models~\cite{Fujikura:2018duw}, Pati-Salam model~\cite{Croon:2018kqn,Huang:2020bbe}, the left-right SU(4) model \cite{Fornal:2018dqn,Fornal:2020ngq} motived by the B physics anomalies, Gorgi-Machacek model~\cite{Zhou:2018zli}, axion or axion-like particle models~\cite{Dev:2019njv, DelleRose:2019pgi, Ghoshal:2020vud}, extra dimensional models~\cite{Yu:2019jlb}, models with charged singlet~\cite{Ahriche:2018rao}, seesaw models~\cite{Brdar:2018num}, models with hidden sectors~\cite{Espinosa:2008kw, Croon:2018erz, Fairbairn:2019xog} and dark matter (DM) models~\cite{Jaeckel:2016jlh, Bird:2016dcv,Beniwal:2018hyi,Bertone:2019irm,Huang:2020mso,Ghosh:2020ipy}, etc. These models reveal that the strong FOPT can produce GW signatures near or above the EW scale~\cite{Dev:2016feu,Weir:2017wfa}.


Among various new physics candidates, except interpreting the EW symmetry breaking by the Higgs mechanism, the minimal left-right symmetric model (LRSM)~\cite{Pati:1974yy, Mohapatra:1974gc, Senjanovic:1975rk} offers an elegant solution to some key fundamental questions in or beyond the SM, such as parity violation/restoration, CP violation, and generation of tiny neutrino masses at the TeV-scale, which are among the focuses of experimental searches of new physics at the high-energy colliders and high-precision experiments. In this work, we examine phase transitions in the LRSM and the resultant features of the corresponding GWs. Compared with the recent and former study \cite{Brdar:2019fur}, the new things of this paper lie in the following aspects:
(i) we have implemented the correct EW vacuum conditions \cite{Chauhan:2019fji} and set $\alpha_2 = 0$ ($\alpha_2$ is a quartic coupling in the scalar potential Eq.~(\ref{eqn:potential})), (ii) we have taken into account more recent LHC experimental bounds, which are collected in Table~\ref{table:bounds} and Fig.~\ref{fig:spectra}, (iii) we have found more general parameter space where the strong FOPT can occur and detectable GWs can be produced, and (iv) we have also explored the complementarity of GW probes of LRSM and the direct searches of the heavy (or light long-lived) particles in the LRSM at the high-energy colliders, and examined how the self couplings of SM Higgs can be affected in the LRSM.

With all the theoretical and experimental limits taken into consideration, it is found that the strong FOPT at the right-handed scale $v_R$ in the LRSM favors relatively small quartic and neutrino Yukawa couplings, which corresponds to relatively light BSM scalars and right-handed neutrinos (RHNs), as seen in Figs.~\ref{figvT}, \ref{fig:random1} and \ref{fig:random2}. The scatter plot in Fig.~\ref{GWpeak} reveals that the phase transition in the LRSM can generate GW signals with the strength of $10^{-17}$ to $10^{-12}$, with a frequency ranging from 0.1 to 10 Hz, which can be probed by the experiments BBO and DECIGO, or even by ALIA and MAGIS. The GW spectra for five benchmark points (BPs) are demonstrated in Fig.~\ref{fig:GWcurves}, which reveals that the GW signal strength and frequency are very sensitive to the value of $\rho_1$. Although some other quartic and neutrino Yukawa couplings are very important for the GW production, the quartic coupling $\rho_1$ plays the most crucial role and it also determines the mass of $SU(2)_R$-breaking scalar $H_3^0$. In the parameter space where it does not mix with other scalars, the scalar $H_3^0$ couples only to the heavy scalars, gauge bosons and RHNs in the LRSM~\cite{Dev:2016dja}, which makes it effectively a singlet-like particle, and thus the experimental limits on it are very weak~\cite{Dev:2016vle,Dev:2017dui}. As presented in Fig.~\ref{fig:complementarity}, the GW probe of $H_3^0$ is largely complementary to the direct searches of $H_3^0$ at the high-energy colliders~\cite{Dev:2016dja} as well as the searches of $H_3^0$ as a long-lived particle (LLP) at the high-energy frontier~\cite{Dev:2016vle,Dev:2017dui}. In addition, in a sizeable region of parameter space, the strong FOPT and GWs are sensitive to a large quartic coupling $\lambda_{hhhh}$ of the SM-like Higgs, which is potentially accessible at a future high-energy muon collider~\cite{Chiesa:2020awd}.

The rest of the paper is organized as follows. In Section~\ref{sec:lrsm} we
briefly review the minimal LRSM and summarize the main existing experimental and theoretical constraints on the BSM particles in this model. Phase transition are explored in Section~\ref{sec:phasetransition}, and the GW production is presented in Section~\ref{sec:GW}. Section~\ref{sec:complementarity} focuses on the complementarity of the GW probes of LRSM and the collider signals of LRSM. After some discussions, we conclude in Section~\ref{sec:conclusion}. For the sake of completeness, the masses and thermal self-energies are collected in Appendix~\ref{appendix:masses}, and the conditions for vacuum stability and correct vacuum are put in Appendix~\ref{appendix:vacuum}.

\section{A brief review of left-right symmetric models}
\label{sec:lrsm}

\subsection{Left-right symmetric model}
\label{sec:model}

The basic idea of LRSMs is to extend the EW sector of $SU(2)_L \times U(1)_Y$ of the SM gauge group to be left-right symmetric, i.e. $SU(2)_L \times SU(2)_R \times U(1)_{B-L}$.  Various LRSMs have been proposed to understand the parity symmetry and CP breaking of the SM, the origin of masses of matters or even DM candidates and the matter-antimatter asymmetry of the universe. The main differences between these LRSMs could be in the gauge structure, the scalar fields, the matter contents, and/or the seesaw mechanisms.

The most popular, or say conventional, LRSM is the version with a Higgs bidoublet $\Phi$, a left-handed triplet $\Delta_L$ and a right-handed triplet $\Delta_R$~\cite{Pati:1974yy, Mohapatra:1974gc, Senjanovic:1975rk}
\begin{eqnarray}
\Phi = \left(\begin{array}{cc}\phi^0_1 & \phi^+_2\\\phi^-_1 & \phi^0_2\end{array}\right) , \;
&& \Delta_L = \left(\begin{array}{cc}\Delta^+_L/\sqrt{2} & \Delta^{++}_L\\\Delta^0_L & -\Delta^+_L/\sqrt{2}\end{array}\right), \;\;\;
\Delta_R = \left(\begin{array}{cc}\Delta^+_R/\sqrt{2} & \Delta^{++}_R\\\Delta^0_R & -\Delta^+_R/\sqrt{2}\end{array}\right).
\label{eq:scalar}
\end{eqnarray}
When the right-handed triplet $\Delta_R$ acquires a vacuum expectation value (VEV) $v_R$, the gauge symmetry $SU(2)_L \times SU(2)_R \times U(1)_{B-L}$ in the LRSM is broken to the SM gauge group  $SU(2)_L \times U(1)_{Y}$. Two triplets $\Delta_L$ and $\Delta_R$ are introduced to give Majorana masses to the active neutrinos and RHNs, respectively, which enables the type-I~\cite{Minkowski:1977sc, Mohapatra:1979ia, Yanagida:1979as, GellMann:1980vs, Glashow:1979nm} and type-II~\cite{Mohapatra:1980yp, Magg:1980ut, Schechter:1980gr, Cheng:1980qt, Lazarides:1980nt}  seesaw mechanisms for the tiny neutrino masses.

The $SU(2)_R \times U(1)_{B-L}$ symmetry can also be broken only by a right-handed doublet $H_R$~\cite{Babu:1988mw,Babu:1989rb}. In this case, heavy vector-like fermions have to be introduced to generate the SM quark and lepton masses via seesaw mechanism (see also~\cite{Mohapatra:2014qva}). There are also LRSM scenarios with inverse seesaw~\cite{Mohapatra:1986aw,Mohapatra:1986bd}, linear seesaw~\cite{Akhmedov:1995ip,Malinsky:2005bi}, or extended seesaw~\cite{Gavela:2009cd,Barry:2011wb,Zhang:2011vh,Dev:2012sg} in the literature. Cold DM is not included in the conventional LRSM (a light RHN can only be a warm DM candidate \cite{Nemevsek:2012cd}), but it is easy to add a fermion or boson multiplet, where the lightest neutral component is naturally stabilized by the residual $Z_2$ symmetry from $U(1)_{B-L}$ breaking~\cite{Heeck:2015qra, Garcia-Cely:2015quu}. Alternatively, based on the gauge group $SU(2)_L \times SU(2)_R \times U(1)_{Y_L} \times U(1)_{Y_R}$ (with $Y_L$ the hypercharge in the SM and $Y_R$ the ``right-handed'' counter partner), heavy RHNs can be the cold DM candidate~\cite{Dev:2016qbd, Dev:2016xcp, Dev:2016qeb}.



In this work, we focus on the minimal LRSM with one bidoublet $\Phi$ and two triplets $\Delta_L$ and $\Delta_R$ in the scalar sector. The most general scalar potential in the LRSM can be written as~\cite{Deshpande:1990ip}
\begin{eqnarray}
\label{eqn:potential}
	{\cal V} &=& -\mu_{1}^{2} \operatorname{Tr}[\Phi^{\dagger} \Phi]-\mu_{2}^{2}\left(\operatorname{Tr}[\tilde{\Phi}
	\Phi^{\dagger}]+\operatorname{Tr}[\tilde{\Phi}^{\dagger} \Phi]\right)-\mu_{3}^{2}\left(\operatorname{Tr}[\Delta_{L}  \Delta_{L}^{\dagger}]+\operatorname{Tr}[\Delta_{R} \Delta_{R}^{\dagger}]\right) \nonumber \\
	  && +\rho_{1}\left(\operatorname{Tr}[\Delta_{L} \Delta_{L}^{\dagger}]^{2}+\operatorname{Tr}[\Delta_{R} \Delta_{R}^{\dagger}]^{2}\right)+\rho_{2}\left(\operatorname{Tr}[\Delta_{L} \Delta_{L}] \operatorname{Tr}[\Delta_{L}^{\dagger} \Delta_{L}^{\dagger}]+\operatorname{Tr}[\Delta_{R} \Delta_{R}] \operatorname{Tr}[\Delta_{R}^{\dagger} \Delta_{R}^{\dagger}]\right) \nonumber \\
	  && +\rho_{3} \operatorname{Tr}[\Delta_{L} \Delta_{L}^{\dagger}] \operatorname{Tr}[\Delta_{R} \Delta_{R}^{\dagger}]+\rho_{4}\left(\operatorname{Tr}[\Delta_{L} \Delta_{L}] \operatorname{Tr}[\Delta_{R}^{\dagger} \Delta_{R}^{\dagger}]+\operatorname{Tr}[\Delta_{L}^{\dagger} \Delta_{L}^{\dagger}] \operatorname{Tr}[\Delta_{R} \Delta_{R}]\right) \nonumber \\
	  && +\lambda_{1} \operatorname{Tr}[\Phi^{\dagger} \Phi]^{2}+\lambda_{2}\left(\operatorname{Tr}[\tilde{\Phi} \Phi^{\dagger}]^{2}+\operatorname{Tr}[\tilde{\Phi}^{\dagger} \Phi]^{2}\right) \nonumber \\
	  && +\lambda_{3} \operatorname{Tr}[\tilde{\Phi} \Phi^{\dagger}] \operatorname{Tr}[\tilde{\Phi}^{\dagger} \Phi]+\lambda_{4} \operatorname{Tr}[\Phi^{\dagger} \Phi]\left(\operatorname{Tr}[\tilde{\Phi} \Phi^{\dagger}]+\operatorname{Tr}[\tilde{\Phi}^{\dagger} \Phi]\right)\nonumber \\
	  && +\alpha_{1} \operatorname{Tr}[\Phi^{\dagger} \Phi]\left(\operatorname{Tr}[\Delta_{L} \Delta_{L}^{\dagger}]+\operatorname{Tr}[\Delta_{R} \Delta_{R}^{\dagger}]\right)+\alpha_{3}\left(\operatorname{Tr}[\Phi \Phi^{\dagger} \Delta_{L} \Delta_{L}^{\dagger}]+\operatorname{Tr}[\Phi^{\dagger} \Phi \Delta_{R} \Delta_{R}^{\dagger}]\right)\nonumber \\
	  && +\left[ \alpha_{2} e^{i \delta} \left(\operatorname{Tr}[\Delta_{L} \Delta_{L}^{\dagger}] \operatorname{Tr}[\tilde{\Phi} \Phi^{\dagger}]+\operatorname{Tr}[\Delta_{R} \Delta_{R}^{\dagger}] \operatorname{Tr}[\tilde{\Phi}^{\dagger} \Phi] \right )+\mathrm{H.c.} \right] \nonumber \\
	  && +\beta_{1}\left(\operatorname{Tr}[\Phi \Delta_{R} \Phi^{\dagger} \Delta_{L}^{\dagger}]+\operatorname{Tr}[\Phi^{\dagger} \Delta_{L} \Phi \Delta_{R}^{\dagger}]\right)+\beta_{2}\left(\operatorname{Tr}[\tilde{\Phi} \Delta_{R} \Phi^{\dagger} \Delta_{L}^{\dagger}]+\operatorname{Tr}[\tilde{\Phi}^{\dagger} \Delta_{L} \Phi \Delta_{R}^{\dagger}]\right) \nonumber \\
	  && +\beta_{3}\left(\operatorname{Tr}[\Phi \Delta_{R} \tilde{\Phi}^{\dagger} \Delta_{L}^{\dagger}]+\operatorname{Tr}[\Phi^{\dagger} \Delta_{L} \tilde{\Phi} \Delta_{R}^{\dagger}]\right), \label{pot}
\end{eqnarray}
where $\tilde{\Phi} = \sigma_2 \Phi^\ast \sigma_2$ (with $\sigma_2$ the second Pauli matrix). Required by left-right symmetry, all the quartic couplings in the potential above are real parameters. The CP violating phase $\delta$ associated with $\alpha_2$ is shown explicitly.

At the zero temperature, the neutral components of the scalar fields can develop non-zero VEVs, i.e.
\begin{equation}
\label{eqn:vev}
	\langle \Phi \rangle=\frac{1}{\sqrt{2}}\left( \begin{array}{cc} \kappa_1 & 0 \\0 & \kappa_2 e^{i \theta_\kappa}\end{array} \right),\quad
	\langle \Delta_{L} \rangle=\frac{1}{\sqrt{2}}\left( \begin{array}{cc} 0 & 0 \\v_L e^{i \theta_L}& 0 \end{array} \right),\quad
	\langle \Delta_{R} \rangle=\frac{1}{\sqrt{2}}\left( \begin{array}{cc} 0 & 0 \\v_R & 0 \end{array} \right) \,,
\end{equation}
where $\theta_\kappa$ and $\theta_L$ are CP violating phases. The two bidoublet VEVs are related to the EW VEV $v_{\rm EW} \simeq (\sqrt 2 G_F)^{-1/2}\simeq 246$ GeV (with $G_F$ the Fermi constant) via $\sqrt{\kappa_1^2+\kappa_2^2}= v_{\rm EW}$. In light of the hierarchy of top and bottom quark masses $m_b \ll m_t$ in the SM, it is a reasonable assumption that $\kappa_2 \ll \kappa_1$~\cite{Deshpande:1990ip}. There are three key energy scales in the LRSM, i.e. the right-handed scale $v_R$, the EW scale $v_{\rm EW}$ and the scale $v_{L}$ which is relevant to tiny active neutrino masses via type-II seesaw. Furthermore, from the first-order derivative of the scalar potential~(\ref{eqn:potential}), $v_L$ is related to the EW and right-handed VEVs via~\cite{Mohapatra:1980yp, Deshpande:1990ip,Kiers:2005gh}
\begin{eqnarray}
v_L = \frac{v_{\rm EW}^2/v_R}{ (1 + \xi^2 ) (2 \rho_1 - \rho_3) }\left[ \beta_1 \xi \cos(\alpha - \theta_L) + \beta_2 \cos \theta_L + \beta_3 \xi^2 \cos(2 \alpha - \theta_L )\right] \,,
\end{eqnarray}
where $\xi = \kappa_2/\kappa_1$. Due to the tiny masses of active neutrinos, it is a good approximation to set $v_L=0$, therefore we will set $\beta_i=0$ so as to simplify our discussions below.

With $v_L = 0$, there are only two energy scales in the LRSM, i.e. the EW scale $v_{\rm EW}$ and the right-handed scale $v_R$. In light of the hierarchy structure $v_{\rm EW} \ll v_R$, a two-step phase transition is supposed to occur in the LRSM. In the early universe, the temperature is so high $T\gg v_R$ that the symmetry $SU(2)_L \times SU(2)_R \times U(1)_{B-L}$ is restored. As the universe keeps expanding, the temperature decreases.
When the temperature is lower than a critical temperature but much higher than EW scale, i.e. $v_{\rm EW} \ll T \sim v_R$, $\Delta_R^0$ develops a non-vanishing VEV and the gauge symmetry $SU(2)_L \times SU(2)_R \times U(1)_{B-L}$ is spontaneously broken to $SU(2)_L \times U(1)_Y$. When the temperature becomes lower than the EW scale $T \sim v_{\rm EW}$, $\Phi_{1,2}^0$ obtain their VEVs and the symmetry is further broken into the electromagnetic (EM) group $U(1)_{\rm EM}$.

After symmetry breaking at the $v_R$ scale, we can rewrite the bidoublet $\Phi$ in terms of two $SU(2)_L$ doublets, i.e. $\Phi = \big( i\sigma_2 H_1^\ast | H_2 \big)$. Then the bidoublet relevant terms in the potential (\ref{eqn:potential}) can be recast in terms of $H_{1,\,2}$:
\begin{align}
\label{eqn:potential2}
{\cal V} (\Phi) \ \supset \ &-m_{11}^2H_1^{\dagger}H_1+m_{22}^2H_2^{\dagger}H_2-m_{12}^2(H_1^{\dagger}H_2+\text{H.c.}) \nonumber \\
&+\lambda_1(H_1^{\dagger}H_1)^2+\lambda_1(H_2^{\dagger}H_2)^2+2\lambda_1H_1^{\dagger}H_1H_2^{\dagger}H_2+4\lambda_3 H_1^{\dagger}H_2 H_2^{\dagger}H_1 \nonumber \\
&+[4\lambda_2(H_1^{\dagger}H_2)^2+2\lambda_4(H_1^{\dagger}H_1+H_2^{\dagger}H_2)H_1^{\dagger}H_2+\text{H.c.}] \,,
\end{align}
where the mass terms are respectively
\begin{eqnarray}
\label{eqn:m11}
m_{11}^2& \ = \ & -\frac{\alpha_{3}}{2} \frac{\kappa_{2}^{2} v_{R}^{2}}{\kappa_{1}^{2}-\kappa_{2}^{2}}+\lambda_{1} v_{\rm EW}^2 +2 \lambda_{4} \kappa_{1} \kappa_{2} \,, \\
\label{eqn:m22}
m_{22}^2& \ = \ & -m_{11}^2+\frac{\alpha_3}{2}v_R^2 \,, \\
\label{eqn:m12}
m_{12}^2& \ = \ & \frac{\alpha_3}{2}\frac{\kappa_1\kappa_2v_R^2}{\kappa_{1}^{2}-\kappa_{2}^{2}}+2(2\lambda_2+\lambda_3)\kappa_1\kappa_2+\lambda_4 v_{\rm EW}^2 \,.
\end{eqnarray}
Although the potential in Eq.~(\ref{eqn:potential2}) seems to be very similar to that in a general 2HDM~\cite{Branco:2011iw}, there are still some obvious differences: In presence of the scale $v_R$, all the states predominately from the heavy doublet $H_2$ are at the $v_R$ scale, and their masses are degenerate at the leading-order, which is clearly distinct from the 2HDMs, where all the scalars in the 2HDMs are at the EW scale, and the BSM scalar masses depend on different quartic couplings~\cite{Branco:2011iw}.

In the LRSM, the BSM particles include the heavy $W_R$ and $Z_R$ bosons, three RHNs $N_i$ (with $i= 1,\, 2,\,3$), neutral CP-even scalar $H_1^0$ and CP-odd $A_1^0$, singly-charged scalar $H_1^\pm$ predominately from the bidoublet $\Phi$, neutral  CP-even scalar $H_2^0$ and CP-odd $A_2^0$, singly-charged scalar $H_2^\pm$ and doubly-charged scalar $H_1^{\pm\pm}$ mostly from the left-handed triplet $\Delta_L$, and the neutral CP-even scalar $H_3^0$ and doubly-charged scalar $H_2^{\pm\pm}$ mostly from the right-handed triplet $\Delta_R$. Thorough studies of the scalar sector of LRSM at future high-energy colliders can be found e.g. in Ref.~\cite{Gunion:1989in, Deshpande:1990ip, Polak:1991vf,  Barenboim:2001vu, Azuelos:2004mwa,  Zhang:2007da, Jung:2008pz,   Bambhaniya:2013wza, Dutta:2014dba, Bambhaniya:2014cia, Bambhaniya:2015ipg, Maiezza:2015lza, Bambhaniya:2015wna, Bonilla:2016fqd, Maiezza:2016ybz, Maiezza:2016bzp, Nemevsek:2016enw, Chakrabortty:2016wkl, Dev:2016dja, Dev:2016vle, Dev:2017dui, Cao:2017rjr, Dev:2018foq, Dev:2018kpa, Chauhan:2019fji}. In this paper, we assume that the gauge coupling $g_R$ for $SU(2)_R$ can be different from the gauge coupling $g_L$ for $SU(2)_L$, which might originate from renormalization group running effects such as in the $D$-parity breaking LRSM versions~\cite{Chang:1983fu}.

\subsection{Theoretical Constraints}
\label{sec:theoretical}

For completeness, we collect all the theoretical constraints on the gauge and scalar sectors of the LRSM in the literature, which will be taken into consideration in the calculations of phase transition and GW production below.

\begin{itemize}
    \item {\it Perturbativity limits}: In some versions of the LRSM, the right-handed gauge coupling $g_R$ can be different from $g_L$~\cite{Chang:1983fu}. As the gauge couplings have the relationship (with $g_{BL}$ the gauge coupling for $U(1)_{B-L}$)
    \begin{eqnarray}
    \frac{1}{e^2} \ = \ \frac{1}{g_L^2} + \frac{1}{g_Y^2} \ = \
    \frac{1}{g_L^2} + \frac{1}{g_R^2} + \frac{1}{g_{BL}^2} \,,
    \end{eqnarray}
    the gauge couplings $g_R$ and $g_{BL}$ can not be either too large or too small if we want them to be perturbative. Renormalization group running these gauge couplings up to a higher energy scale put more stringent limits on them. Perturbativity up to the GUT scale requires that the ratio $r_g \equiv g_R / g_L$ to satisfy~\cite{Chauhan:2018uuy}\footnote{Note that the perturbativity limits in Ref.~\cite{Chauhan:2018uuy} are on the LRSM without the left-handed triplet $\Delta_L$  at the TeV-scale. In presence of $\Delta_L$ at the TeV-scale, the perturbativity limits should be to some extent different. As an approximation we will adopt the limits from Ref.~\cite{Chauhan:2018uuy}.}
    \begin{eqnarray}
    0.65 < r_g < 1.60 \,.
    \end{eqnarray}
    Furthermore, as the masses $\sqrt{\alpha_3/2} v_R$ of $H_1^0$, $A_1^0$ and $H_1^\pm$ (cf. Table~\ref{tab:mass} in Appendix~\ref{appendix:masses}) are severely constrained by the neutral meson mixings (see Section~\ref{sec:theoretical} and Table~\ref{table:bounds}), perturbativity also implies an lower bound on the $v_R$ scale~\cite{Chauhan:2018uuy}:
    \begin{eqnarray}
    v_R \gtrsim 10\, {\rm TeV} \,.
    \end{eqnarray}
    For $v_R$ below this value, $\alpha_3$ is so large that all the quartic and gauge couplings will hit the Landau pole very quickly before reaching the GUT or Planck scale~\cite{Rothstein:1990qx, Chakrabortty:2013zja, Chakrabortty:2016wkl, Maiezza:2016ybz}.

    \item {\it Unitarity conditions}: The parameters in the potential (\ref{pot}) should satisfy the unitarity conditions~\cite{Chakrabortty:2016wkl} when we consider the scattering amplitudes of the scalar fields at the high-energy scale $\sqrt{s} \gg \mu^2_i$ (for simplicity we neglect here the effects of all the scalar masses). In other words, the partial wave amplitudes should not violate the bound of unitarity so as to guarantee that the probability is conserved. The tree-level unitarity conditions turn out to be~\cite{Chakrabortty:2016wkl}
\begin{eqnarray}
&& \lambda_{1,\,4} < \frac{4 \pi}{3} \,, \quad
\lambda_1 + 4 \lambda_2 + 2 \lambda_3  < 4 \pi \,, \quad
\lambda_1 - 4 \lambda_2 + 2 \lambda_3  <  \frac{4 \pi}{3} \,, \nonumber \\
&& \rho_1 < \frac{4 \pi}{3} \,, \quad
\rho_1 + \rho_2  < 2 \pi \,, \quad
 \rho_{2,\,4}  <  2 \sqrt{2} \pi \,, \quad
\rho_3  < 8 \pi \,, \nonumber \\
&& \alpha_1 <  8 \pi \,, \quad
\alpha_2 <  4 \pi \,, \quad
\alpha_1 + \alpha_3 <  8 \pi \,.
\end{eqnarray}

\item {\it Vacuum stability conditions}: The vacuum stability conditions require that \cite{Chakrabortty:2013zja, Chakrabortty:2013mha, Chakrabortty:2016wkl} (see also~\cite{Kannike:2016fmd})
\begin{eqnarray}
\lambda_1  >  0\,\,, \quad
\rho_1  >  0\,\,, \quad
\rho_1 + \rho_2  >  0\,\,, \quad
\rho_1 + 2 \rho_2  >  0\,\,.
\end{eqnarray}

\item {\it Correct vacuum criteria}: After the spontaneous symmetry breaking, all the scalar fields have to form some specific structure in the phase space such that we reside in the correct vacuum, i.e. the vacuum with the lowest VEV in the potential~\cite{Dev:2018foq, Chauhan:2019fji}. For completeness, the correct vacuum criteria have been collected in Appendix~\ref{appendix:vacuum}, which are obtained with the assumption $\alpha_2=0$. Therefore, we will set $\alpha_2=0$ throughout this paper.

In the limit of $\kappa_2 \ll \kappa_1 \ll v_R$, in Eq.~(\ref{eqn:m22}),
the quadratic coefficient of $H_2$ term is proportional to $\alpha_{3}v_R^2/2$, thus the heavy doublet scalars $H_1^0, A_1^0, H_2^\pm$ will obtain a mass of $ \sqrt{\alpha_3/2}  v_R$ at the leading-order. To get the correct EW vacuum, a necessary condition is $m_{11}^2>0$, i.e.
\begin{equation}
-\frac{\alpha_{3}}{2} \frac{\kappa_{2}^{2} v_{R}^{2}}{\kappa_{1}^{2}-\kappa_{2}^{2}}+\lambda_{1} v_{\rm EW}^2 +2 \lambda_{4} \kappa_{1} \kappa_{2}>0.
\end{equation}
This yields an upper bound of $\xi$. Approximately, we have
\begin{eqnarray}
	\label{eqn:xi}
\xi & \ \lesssim \ & \frac{\sqrt{\lambda_1} v_{\rm EW}}{M_{H_1^0}} \nonumber \\
& \ \lesssim \ & 8.9 \times 10^{-3}
\left( \frac{\lambda_1}{0.13} \right)^{1/2}
\left( \frac{m_{H_1^0}}{10\,{\rm TeV}} \right)^{-1} \,.
\end{eqnarray}


\end{itemize}


\subsection{Experimental constraints}
\label{sec:experimental}

\begin{figure}[!t]
	\centering
	\includegraphics[width=0.75\textwidth]{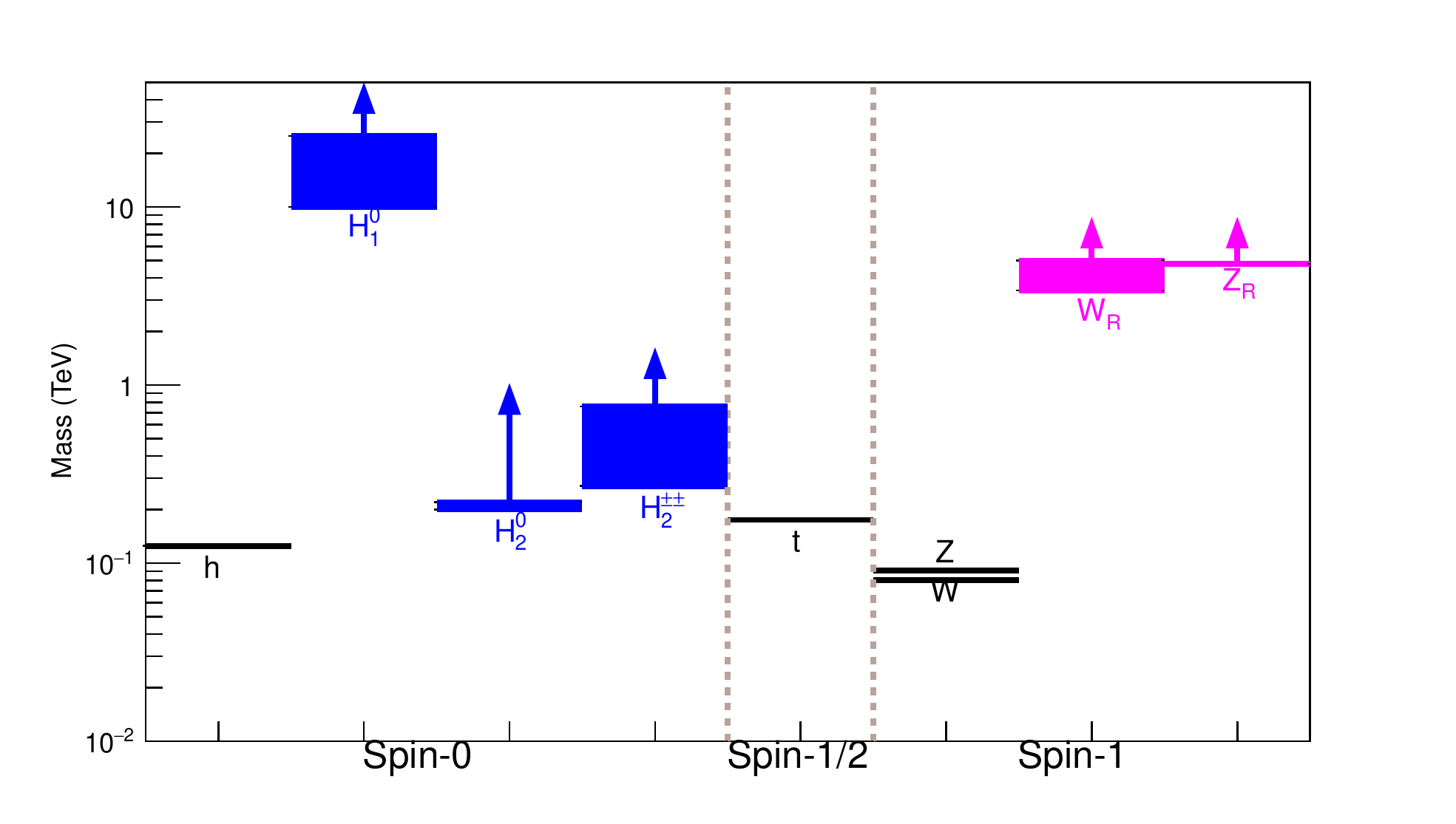}
	\caption{Experimental limits on the scalars and gauge bosons in Table~\ref{table:bounds}, indicated by the blue and pink arrows, with the heights of the horizontal lines denoting the ranges of experimental limits. The horizontal black lines are the masses of SM Higgs $h$, top quark $t$, and $W$, $Z$ bosons.
	\label{fig:spectra} }
\end{figure}

All the current LHC limits on the BSM particles in the LRSM are collected in Table~\ref{table:bounds} and also depicted in Fig.~\ref{fig:spectra}. Here are more details:
\begin{itemize}
    \item At the LHC, the $W_R$ boson in the LRSM can be produced via the right-handed charged quark currents. After its production, it can decay predominately into two quark jets (including the $\bar{t}b$ channel) and RHNs plus a charged lepton, i.e. $W_R \to jj,\, \bar{t}b,\, N_i^{(\ast)} \ell_\alpha$ (with $\alpha = e,\,\mu,\,\tau$). If the RHNs are lighter than the $W_R$ boson, as a result of the Majorana nature of RHNs, the same-sign dilepton plus jets $W_R \to N \ell \to \ell_\alpha \ell_\beta j j$ constitute a smoking-gun signal of the $W_R$ boson~\cite{Keung:1983uu}. Assuming $g_R = g_L$, the current most stringent LHC data require that the $W_R$ mass $m_{W_R}> (3.8 - 5)$ TeV for a RHN mass $100 \, {\rm GeV} < m_{N} < 1.8$ TeV~\cite{Aaboud:2018spl, Aaboud:2019wfg}. The dijet~\cite{Aad:2019hjw,Sirunyan:2019vgj} and $\bar{t}b$~\cite{Sirunyan:2017ukk,Sirunyan:2017vkm} limits are relatively weaker, which are respectively 4 TeV and 3.4 TeV. The strongest $W_R$ limit of $(3.8 - 5)$ TeV is presented in Fig.~\ref{fig:spectra}.

    \item The most stringent limits on the $Z_R$ boson is from  the dilepton data $pp \to Z_R \to \ell^+ \ell^-$. The current dilepton limit on a sequential $Z'$ boson is 5.1 TeV~\cite{Aad:2019fac}. Following e.g. Ref.~\cite{Chauhan:2018uuy}, one can rescale the production cross section times branching fraction $\sigma (pp \to Z' \to \ell^+ \ell^-)$ for the sequential $Z'$ model, which leads to the LHC dilepton limit of 4.82 TeV on the $Z_R$ boson in the LRSM. This is shown in Fig.~\ref{fig:spectra} as the $Z_R$ limit. There are also dijet searches of the $Z'$ boson, however the corresponding limits are relatively weaker~\cite{Aad:2019hjw,Sirunyan:2019vgj}.




\item At the leading order, the scalars $H_2^0$, $A_2^0$, $H_2^\pm$ and  $H_1^{\pm\pm}$ from the left-handed triplet $\Delta_L$ have the same mass~\cite{Zhang:2007da} (see Table~\ref{tab:mass}).
      The doubly-charged scalar $H_1^{\pm\pm}$ can decay into either same-sign dilepton or same-sign $W$ bosons, i.e. $H_1^{\pm\pm} \to \ell_\alpha^\pm \ell_\beta^\pm,\, W^\pm W^\pm$, which constitute the most promising channels to probe $\Delta_L$ at the LHC, and the branching fractions ${\rm BR} (H_1^{\pm\pm} \to \ell_\alpha^\pm \ell_\beta^\pm)$ and  ${\rm BR} (H_1^{\pm\pm} \to  W^\pm W^\pm)$ depend on the Yukawa coupling $f_L$ and the left-handed triplet VEV $v_L$. Assuming $H_1^{\pm\pm}$ decays predominately into electrons and muons, the current LHC limits are around 770 to 870 GeV, depending on the flavor structure~\cite{Aaboud:2017qph}. In the di-tauon channel $H_1^{\pm\pm} \to \tau^\pm \tau^\pm$, the LHC limit is relatively weaker, i.e. 535 GeV~\cite{CMS:2017pet}.\footnote{
      As the singly-charged scalar $H_2^{\pm}$ and doubly-charged scalar $H_1^{\pm\pm}$ are mass degenerate at the leading order in the LRSM, here we have adopted the combined LHC limit from the pair production $pp \to H_1^{++} H_1^{--}$ and the associate production $pp \to H_1^{\pm\pm} H_2^{\mp}$. In these two channels, the separate channels are respectively 396 GeV and 479 GeV~\cite{CMS:2017pet}.}  If the doubly-charged scalar $H_1^{\pm\pm}$ decays predominately into same-sign $W$ bosons, the LHC limits are much weaker, around 200 to 220 GeV~\cite{Aaboud:2018qcu}. There are also some searches of singly-charged scalar $H_2^{\pm} \to \tau^\pm \nu$ at the LHC~\cite{Aaboud:2016dig, Aaboud:2018gjj, Sirunyan:2019hkq}. However these searches assume $H_2^\pm$ is produced from its interaction with top and bottom quarks, therefore these limits are not applicable to $H_2^{\pm}$ in the LRSM which does not couple directly to the SM quarks. The strongest same-sign dilepton limits of $(530 - 870)$ GeV on $H_1^{\pm\pm}$ (and also on other scalars from $\Delta_L$) is shown in Fig.~\ref{fig:spectra}.

  \item  As the $W_R$ boson is very heavy, the TeV-scale right-handed doubly-charged scalar $H_2^{\pm\pm}$ decays only into same-sign dileptons. The couplings of $H_2^{\pm\pm}$ to the photon and $Z$ boson have opposite signs, therefore the production cross section of $H_2^{\pm\pm}$ at the LHC is smaller than that for the left-handed  doubly-charged scalar $H_1^{\pm\pm}$. Rescaling the LHC13 cross section of $H_1^{\pm\pm}$ by a factor of 1/2.4,
  The same-sign dilepton limits on $H_2^{\pm\pm}$ turn out to be 271 to 760 GeV for all the six combinations $ee,\, e\mu,\, \mu\mu,\, e\tau,\, \mu\tau,\, \tau\tau$ of lepton flavors,  which is presented in Fig.~\ref{fig:spectra}.

    \item The scalars $H_1^0$, $A_1^0$ and $H_1^\pm$ from the bidoublet $\Phi$ are degenerate in mass at the leading order. $H_1^0$ and $A_1^0$ has tree-level flavor-changing neutral-current (FCNC) couplings to the SM quarks, and contribute to $K-\overline{K}$, $B_d-\overline{B}_d$ and $B_s-\overline{B}_s$ mixings significantly. As a result, their masses are required to be at least  $(10 - 25)$ TeV, depending on the nature of left-right symmetry (either generalized parity or generalized charge conjugation), the hadronic uncertainties~\cite{Ecker:1983uh, Zhang:2007da, Maiezza:2010ic, Bertolini:2014sua} and the potentially large QCD corrections~\cite{Bernard:2015boz}. The stringent FCNC limits on the heavy bidoublet scalars is shown in Fig.~\ref{fig:spectra}.

   \item The neutral scalar $H_3^0$ from the right-handed triplet $\Delta_R$ is hadrophobic, i.e. it does not couples directly to the SM quarks in the Lagrangian. It can be produced at the LHC and future higher energy colliders either in the scalar portal through coupling to the SM Higgs (and  the heavy scalars $H_1^0$ and $A_1^0$), or in the gauge portal via coupling to the $W_R$ and $Z_R$ bosons. Therefore the {\it direct} LHC limits are very weak~\cite{Dev:2016vle,Dev:2017dui}. However, when it is sufficiently light, say at the GeV-scale, $H_3^0$ can be produced from (invisible) decay of the SM Higgs or even from the meson decays~\cite{Dev:2016vle,Dev:2017dui}. More details can be found in Section~\ref{sec:H3}.

   \item The RHNs in the LRSM can be either very light, e.g. at the keV scale to be a warm DM~\cite{Nemevsek:2012cd} candidate, or very heavy at the $v_R$ scale, and there are almost no laboratory limits on their masses, although their mixings with the active neutrinos are tightly constrained in some regions of the parameter space~\cite{Bolton:2019pcu}. For simplicity, in the following sections we will set the masses of RHNs to be free parameters and neglect their mixings with the active neutrinos.
\end{itemize}

\begin{center}
\begin{table}
  \begin{center}
  \caption{Current most stringent experimental limits on the masses of $W_R$, $Z_R$, $H_1^{\pm\pm}$, $H_2^{\pm\pm}$, and $H_1^0$, $A_1^0$ in the LRSM. The particles in parentheses are mass degenerate with them, if there is any. See text for more details.  \vspace{5pt}
  \label{table:bounds}}
  \begin{tabular}{c|ccc}
  \hline\hline
  Particle   &   Channel   &   Lower Limit   &    References  \\
  \hline
   \multirow{3}{*}{$W_R$} & $\ell\ell jj$   &   $3.8 - 5.0$ TeV   &    \cite{Aaboud:2019wfg, Aaboud:2018spl}  \\
             &   $jj$      &   $4.0$ TeV   &    \cite{Aad:2019hjw,Sirunyan:2019vgj}  \\
             &   $t\bar{b}$   &   $3.4$ TeV   &    \cite{Sirunyan:2017ukk,Sirunyan:2017vkm}  \\  \hline
  $Z_R$        &    $\ell^+\ell^-$  &  $4.8$ TeV &  \cite{Aad:2019fac} \\ \hline
  $H_1^{\pm\pm}$   &   $\ell_\alpha^{\pm} \ell_\beta^{\pm}$    &   $535 - 870$ GeV &   \cite{Aaboud:2017qph, CMS:2017pet}  \\
  ($H_2^0$, $A_2^0$, $H_2^\pm$)   &   $W^{\pm}W^{\pm}$  &  $200 - 220$ GeV   &   \cite{Aaboud:2018qcu}  \\  \hline
  $H_2^{\pm\pm}$   &   $\ell_\alpha^{\pm} \ell_\beta^{\pm}$    &  $271 - 760$ GeV    &   \cite{Aaboud:2017qph}  \\ \hline
  $H_1^0$, $A_1^0$ ($H_1^{\pm}$) & meson mixing & 10 - 25 TeV &  \cite{Ecker:1983uh, Zhang:2007da, Maiezza:2010ic, Bertolini:2014sua}  \\ 
  \hline
  \hline
  \end{tabular}
  \end{center}
\end{table}
\end{center}

To be complete, the masses of 100 GeV scale SM particles, i.e. the SM Higgs $h$, the top quark $t$ and the $W$ and $Z$ bosons, are depicted in Fig.~\ref{fig:spectra} as horizontal black lines. See Fig.~\ref{fig:cmbspectra} for complementarity of GW prospects of the BSM particle masses and the current experimental limit.


\section{Phase transition in LRSM}
\label{sec:phasetransition}

\subsection{One-loop effective potential}

To study phase transitions in the LRSM, we consider the effective potential at finite temperature, which includes contributions of the one-loop corrections and daisy resummations. Renormalized in the $\overline{\text{MS}}$ scheme, the effective potential can be cast into the following form \cite{Basler:2018cwe}
\begin{eqnarray}
     \label{eqn:Veff}
     {\cal V}_{\rm eff}(\phi_i,v) & \  = \ & V_0(\phi_i,v)+V_1^{T=0}(\phi_i,v)+V_1^{T \neq 0}(\phi_i,v)+V_D(\phi_i,v) \nonumber \\
     & \ = \ & V_0(\phi_i,v)+\frac{1}{64 \pi^2}\sum_{i} g_i m_i^4(\phi_i,v) \left(\log\frac{m_i^2(\phi_i,v)}{\mu^2}-C_i \right) \nonumber \\
     &&  +\frac{T^4}{2 \pi^2}\sum_{i} g_i J_{\pm}\left(\frac{m_i^2(\phi_i,v)}{T^2} \right) \nonumber \\
     &&  -\frac{T}{12\pi}\sum_{i={\rm bosons}} \left[\left( m_i^2(\phi_i,v)+\Pi_i(T)\right)^{3/2} - \left(m_i^2(\phi_i,v) \right)^{3/2} \right],
\end{eqnarray}
where $V_0(\phi_i,v)$  is the tree-level potential, $V_1^{T=0}$ is the Coleman-Weinberg one-loop effective potential~\cite{Coleman:1973jx}, and $V_1^{T\neq 0}$ and $V_D$ are the thermal contributions at finite temperature. The $V_1^{T\neq0}$ term includes only the one-loop contributions, and $V_D$ denotes the high-order contributions from daisy diagrams. In Eq.~(\ref{eqn:Veff}) the sum runs over all the particles in the model. The scalar mass matrices $m_i^2(\kappa_i,v_R)$ in the LRSM can be found in Ref.~\cite{Deshpande:1990ip}, and the corresponding thermal self-energies  $\Pi_i(T)$ are  provided in Appendix~\ref{appendix:masses}. As for the fermions, we consider only the third generation quarks and three RHNs. In the LRSM their masses are respectively
\begin{eqnarray}
m_t=\frac{1}{\sqrt{2}}(y_t\kappa_1+y_b\kappa_2) \,, \quad m_b=\frac{1}{\sqrt{2}}(y_b\kappa_1+y_t\kappa_2) \,, \quad
M_N=\sqrt{2}y_Nv_R \,,
\end{eqnarray}
with $y_{t,\,b}$ the Yukawa couplings for top and bottom quarks in the SM, $M_N$ the RHN masses and $y_N$ the corresponding Yukawa coupling. In the following study, for the sake of simplicity, we will assume three RHNs are mass degenerate and does not have any mixings among them. The degrees of freedom $g_i$ and constants $C_i$ in Eq.~(\ref{eqn:Veff}) are given by
    \begin{equation}
    	\begin{aligned}
    	(g_i,C_i)=\left\{\begin{array}{ll}
    	(1,\frac{3}{2}), & \mbox{for scalars} \,,\\
    	(-2\lambda,\frac{3}{2}),  & \mbox{for fermions} \,, \\
    	(3,\frac{5}{6}),   & \mbox{for gauge bosons} \,,
    	\end{array}\right.
    	\end{aligned}
    \end{equation}
with $\lambda=1 \, (2)$ for Weyl (Dirac) fermions, and the functions $J_{-} (J_{+})$ for bosons (fermions) are defined as
\begin{equation}
    J_{\pm}(x^2)= \int_{0}^{\infty}dk  k^2 \log \left(1 \pm e^{-\sqrt{x^2+k^2}} \right) \,.
\end{equation}
In the limit of small $x^2 = m^2/T^2$, we can use the approximations~\cite{Basler:2018cwe}:
\begin{eqnarray}
 J_{+}\left(x^{2}\right) & \ = \ & \frac{7 \pi^{4}}{360}-\frac{\pi^{2}}{24} x^{2}-\frac{1}{32} x^{4}\log \frac{x^{2}}{a_F}+\mathcal{O}(x^4) \,, \\
 J_{-}\left(x^{2}\right) & \ = \ & -\frac{\pi^{4}}{45}+\frac{\pi^{2}}{12} x^{2}-\frac{\pi}{6}\left(x^{2}\right)^{3 / 2}-\frac{1}{32} x^{4}\log \frac{x^{2}}{a_B}+\mathcal{O}(x^4) \,,
\end{eqnarray}
where
\begin{eqnarray}
a_F \ =  \ \pi^2 e^{3/2 -2\gamma_E} \,, \quad
a_B \ = \ 16\pi^2 e^{ 3/2 -2\gamma_E }\,.
\end{eqnarray}

In this paper we focus on the phase transition at the $v_R$ scale, thus as an approximation all the effects of SM components on the symmetry breaking $ SU(2)_R \times U(1)_{B-L} \to U(1)_Y$ can be neglected. Neglecting the daisy contributions, the effective potential ${\cal V}_{\rm eff}$ can be written down explicitly in the following form~\cite{Cohen:1993nk}:
\begin{equation}
\begin{aligned}
V_{eff}( v, \Pi_i =0) \ \simeq \  D \, (T^2 - T_0^2)\, v^2 - E \, T \, v^3 + \frac{\rho_T}{4} \,  v^4\,,
\end{aligned}
\label{vpara}
\end{equation}
where $D$, $T_0$, $E$ and $\rho_T$ can be expressed by the model parameters as
\begin{eqnarray}
\label{eqn:D}
\label{eqn:D}
D & \ = \ & \frac{1}{8 v_R^2}\left(M_{Z_R}^2+2M_{W_R}^2+M_{N}^2\right)+D_H \,, \\
\label{eqn:T0}
T_0^2 & = & \frac{M_{H_3^0}^2}{4D}+T^2_H \,, \\
\label{eqn:E}
E & \ = \ & \frac{ M_{Z_R}^3+2M_{W_R}^3}{4\pi v_R^3}+E_H \,, \\
\label{eqn:rhoT}
\rho_T & \ = \ & \rho_1-\frac{3\left(M_{Z_R}^4+2M_{W_R}^4\right)}{16\pi^2 v_R^4}\left(\frac{5}{6}+\log\frac{\mu^2}{a_BT^2}\right)  \nonumber \\
&& +\frac{6M_{N}^4}{16\pi^2 v_R^4}\left(\frac{3}{2}+\log\frac{\mu^2}{a_FT^2}\right)+\rho_H \,,
\label{eqply1}
\end{eqnarray}
where $M_X$ is the mass for the particle $X$, and $\mu$ is the renormalization scale. Since there are lots of scalars in the LRSM, we deliberately separate their contributions from the vector bosons and RHNs. The contributions of scalars for each of the terms in Eq.~(\ref{eqn:D}) to (\ref{eqn:rhoT}) can be written in terms of the scalar masses via
\begin{eqnarray}
\label{eqn:DH}
D_H & \ = \ & \frac{1}{24 v_R^2}\left( 4M^2_{H_1^0}  +6M_{H_2^0}^2 +7M_{H_3^0}^2 +2M_{H_2^{\pm\pm}}^2     \right) \,, \\
\label{eqn:TH}
T^2_H & \ = \ & \frac{M_{H_3^0}^2}{D}\frac{ 6M_{H_2^0}^2 + 7M_{H_3^0}^2+2M_{H_2^{\pm\pm}}^2}{64\pi^2v_R^2}\left(\frac{3}{2}+\log \frac{\mu^2}{a_BT^2} \right) \,, \\
\label{eqn:EH}
E_H & \ = \ & \frac{1}{16\pi v_R^3} \left\{
\frac{16}{3} M_{H_1^0}^3 +
\sqrt2 M_{H_3^0}^3 \left( 1 - r_v \right)^{3/2} +
\sqrt6 M_{H_3^0}^3 \left( 1 - \frac13 r_v \right)^{3/2} \right. \nonumber \\
&& + \left. 2\sqrt2 \left[ M_{H_3^0}^2 \left( 1 - r_v \right) + 2 M_{A_2^0}^2 \right]^{3/2}
+ \frac{2\sqrt2}{3} \left[ M_{H_3^0}^2 \left( 1 - r_v \right) + 2 M_{H_2^{\pm\pm}}^2 \right]^{3/2} \right\} \,, \\
\label{eqn:rhoH}
\rho_H & \ = \ & -\frac{ 4M_{H_1^0}^4 + 6M_{H_2^0}^4 + 5M_{H_3^0}^4 + 2M_{H_2^{\pm\pm}}^4  +6M_{H_2^0}^2M_{H_3^0}^2 + 2M_{H_3^0}^2M_{H_2^{\pm\pm}}^2}{16\pi^2 v_R^4} \nonumber \\
&& \quad\times\left(\frac{3}{2}+\log\frac{\mu^2}{a_BT^2}\right) \,,
\end{eqnarray}
where we have defined $r_v \equiv v_R^2/v^2$. It should be pointed out that all the masses in Eqs.~(\ref{eqn:DH}) to (\ref{eqn:rhoH}) depend upon the right-handed VEV $v_R$ instead of $v$. It is observed that the RHNs can also contribute to the symmetry breaking $SU(2)_R \times U(1)_{B-L} \to U(1)_{Y} $ via affecting the parameters $D$, $T_0$ and $\rho_T$, while the parameter $E$ receives only contributions from the scalars and gauge bosons.

As seen in Eqs.~(\ref{eqn:rhoT}) and (\ref{eqn:rhoH}), the parameter $\rho_T$ receive not only tree-level contribution from the quartic coupling $\rho_1$ which corresponds to the  $H_3^0$ mass via $\rho_1 \simeq M_{H_3^0}^2/2v_R^2$ (see Table~\ref{tab:mass}), but also loop-level contributions from the heavy scalars, gauge bosons and RHNs in the LRSM. In particular, when the quartic coupling $\rho_1$ is small, or equivalently the scalar $H_3^0$ is much smaller than the $v_R$ scale, which is the parameter space of interest for phase transition and GW production in the LRSM (cf. Figs.~\ref{figvT}, \ref{fig:random1} and \ref{fig:GWcurves}), the loop-level contributions in Eq.~(\ref{eqn:rhoT}) might dominate $\rho_T$. Furthermore, $\rho_T$ depends also on the gauge coupling $g_R$ via the heavy gauge boson masses  $M_{W_R}$ and $M_{Z_R}$.

To have strong FOPT, the cubic terms proportional to $- E T v^3$ are crucial. In the limit of $E \to 0$, the phase transition is of second-order. In the SM, the effective coefficient $E$ of $\phi^3$ term is dominated by the gauge boson contributions, while in the LRSM, it receives contributions from both the scalars and gauge bosons, As a result of the large degree of freedom in the scalar sector of LRSM, it is remarkable that the scalar contributions to $E$ can even be much larger. The order parameter describing the FOPT is given by $v_c/T_c$, where $v_c$ is the non-vanishing location of the minimum at the critical temperature $T_c$ at which the effective potential ${\cal V}_{\rm eff}$ has two degenerate minima. In the EW baryogenesis~\cite{Kuzmin:1985mm, Shaposhnikov:1986jp, Shaposhnikov:1987tw}, to avoid the washout effects in the broken phase within the bubble wall,  a strong FOPT is typically required to satisfy the following condition
\begin{eqnarray}
\label{eqn:vcTc}
\frac{v_c}{T_c} = \frac{2 E}{\rho_T} \geq 1 \,.
\end{eqnarray}

\subsection{Strong first-order phase transition at the $v_R$ scale}

The effective potential (\ref{eqn:Veff}) is a function of temperature $T$. Meanwhile, the minima of the effective potential vary when the temperature changes. In order to find the quantity $v_c/T_c$ which measures the strength of FOPT, we need to find both the critical temperature $T_c$ and the critical VEV $v_c$.\footnote{There might be some theoretical uncertainties in perturbative calculations of FOPTs and resultant GWs, which can be found, e.g. in Ref.~\cite{Croon:2020cgk}. }
In term of the parametrization given in Eq.~(\ref{vpara}), the critical temperature can be approximately expressed as
\bea
T_c^2 \simeq T_0^2  \frac{\rho_T D }{\rho_T D - E^2}\,.
\eea
Thus it is clear that $T_c \sim v_R \gg v_{\rm EW}$. Therefore, it is justified to neglect the contributions of SM particles to the phase transition at the right-handed scale $v_R$, since their masses $m_{\rm SM}$ are at most close to $v_{\rm EW}$ and their contributions are suppressed due to their tiny couplings to the right-handed triplet.

\begin{figure}[!t]
\includegraphics[height=0.4\linewidth]{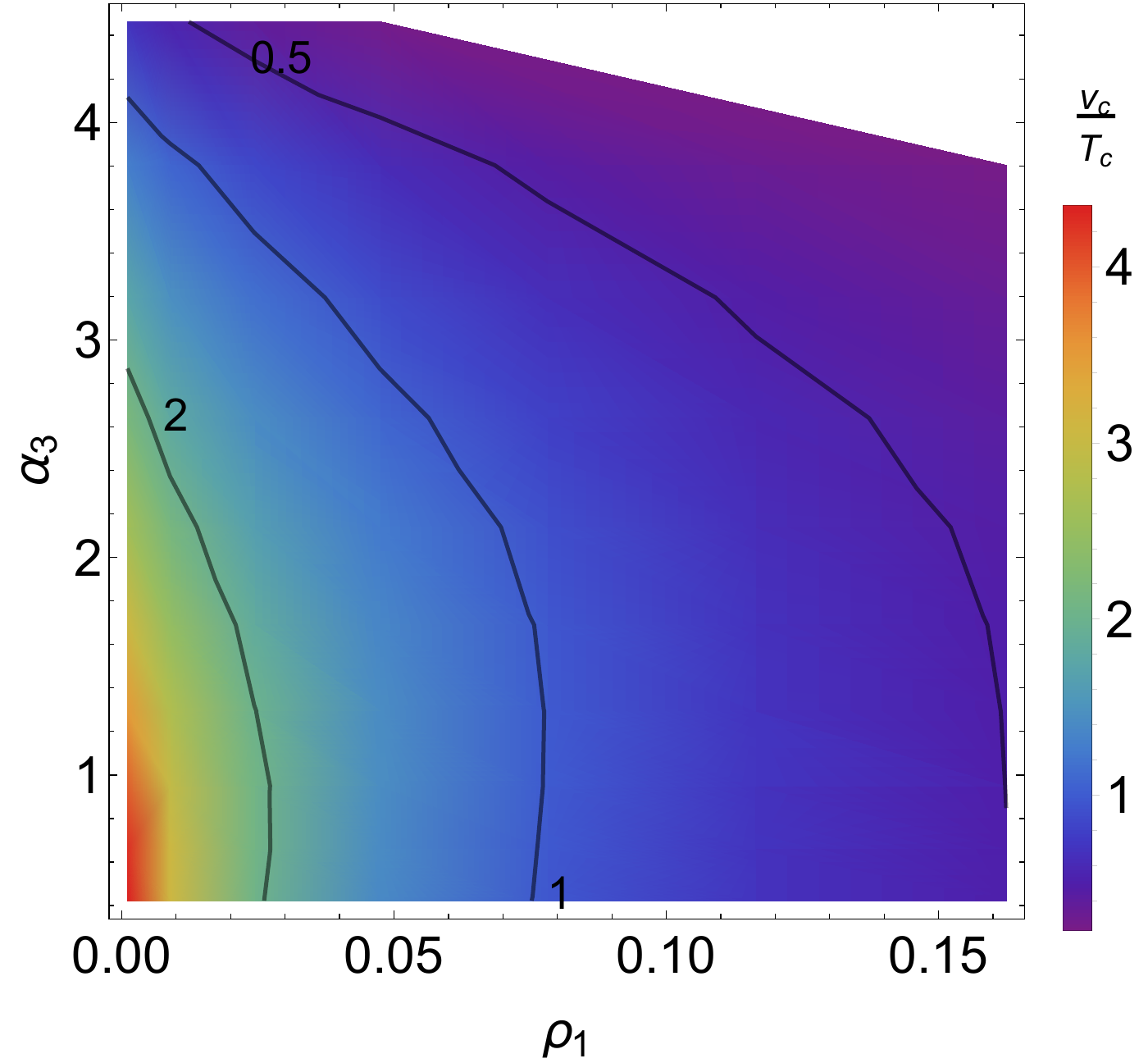} \hspace{0.45cm}
\includegraphics[height=0.415\linewidth]{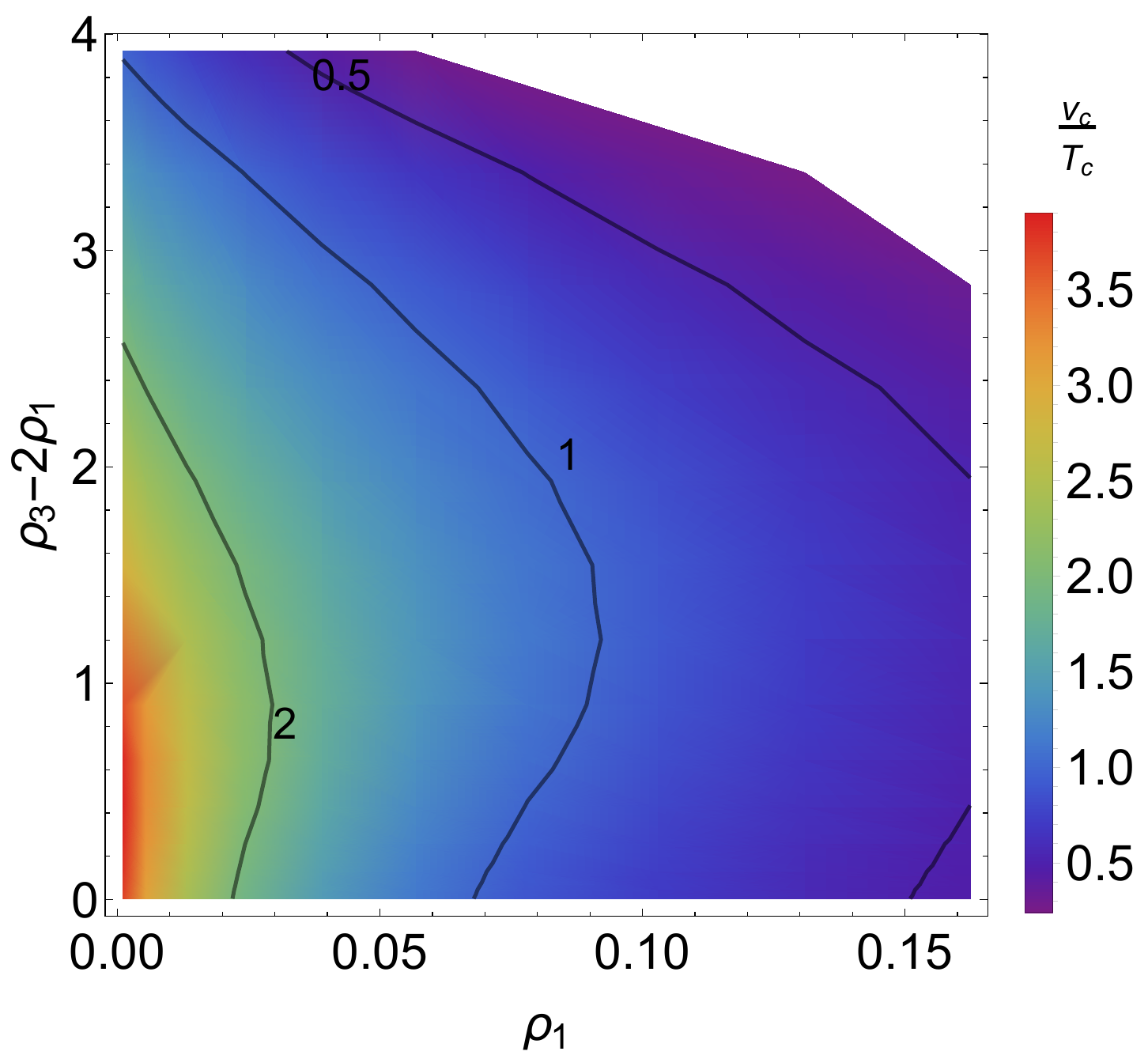} \\
\includegraphics[height=0.4\linewidth]{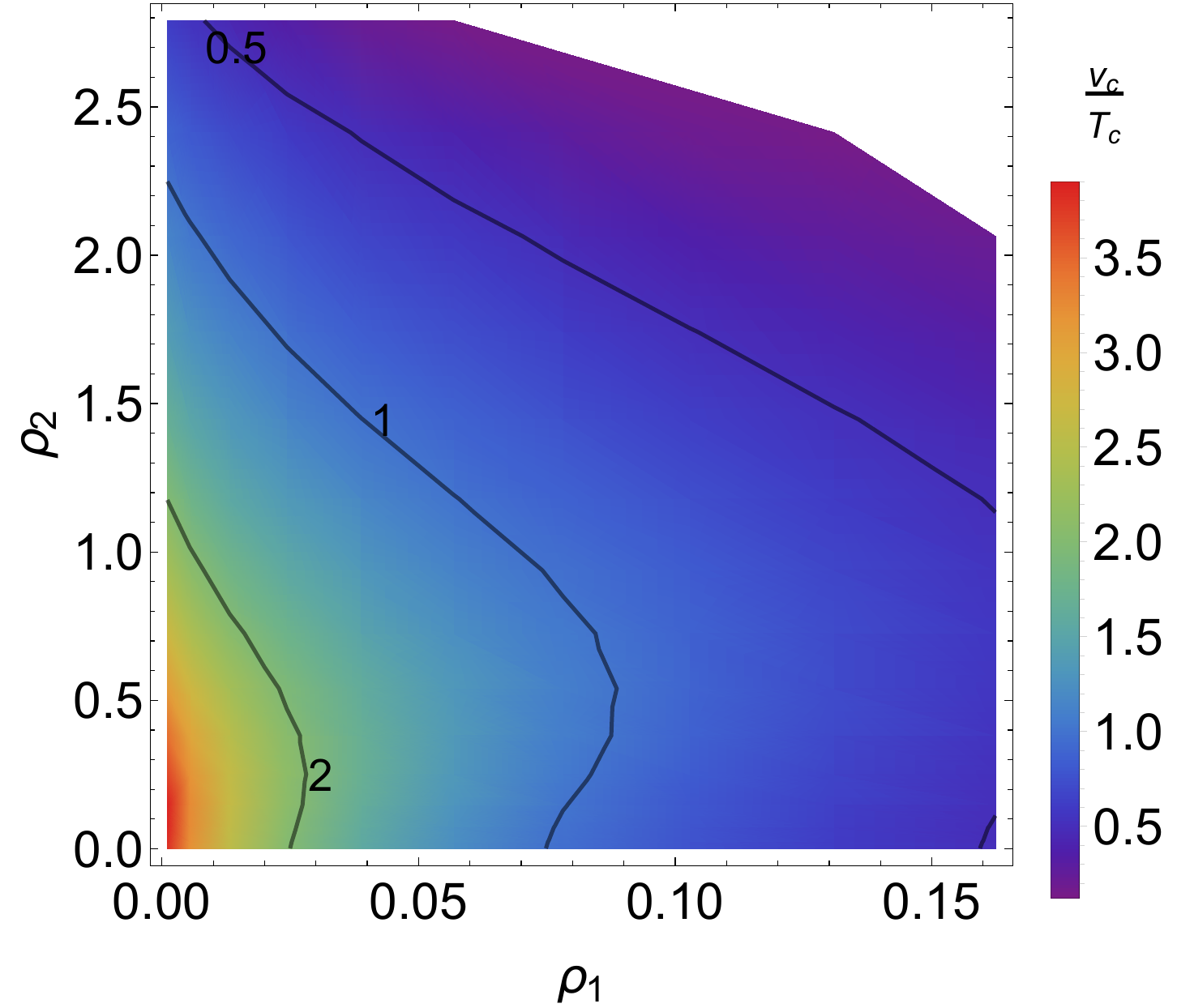} \hspace{-0.3cm}
\includegraphics[height=0.41\linewidth]{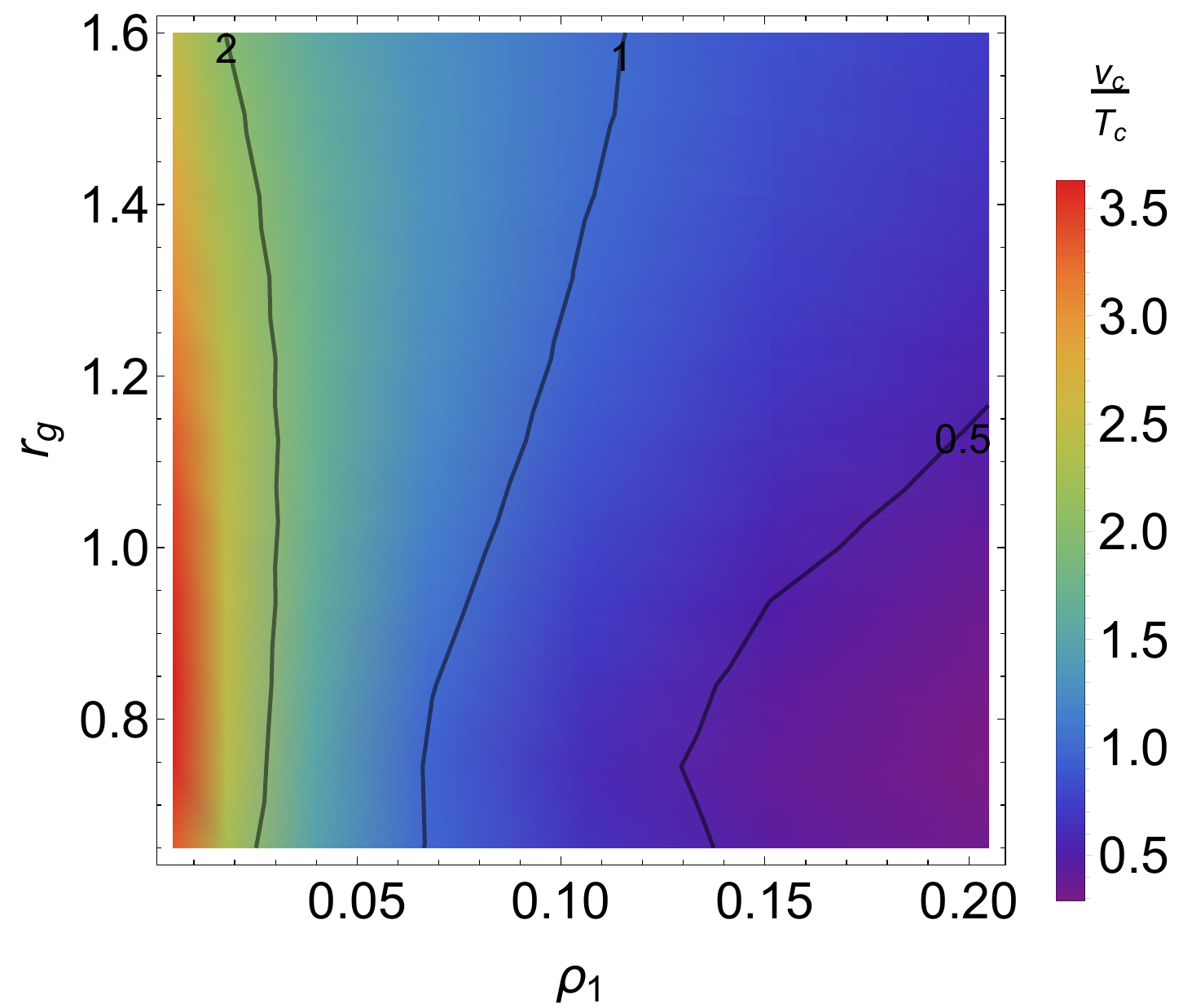}
\caption{$v_c/T_c$ at critical temperature in the plane of $\rho_1$ versus $\alpha_3$ (upper left), $\rho_1$ versus $\rho_3 -2\rho_1$ (upper right), $\rho_1$ versus $\rho_2$ (lower left) and $\rho_1$ versus $r_g$ (lower right). The color indicates the value of $v_c/T_c$. In all the panel the other parameters are fixed in the way that their corresponding scalar masses are set to be the $W_R$ mass.
\label{figvT}}
\end{figure}

For given $v_R$ and heavy particle masses in the LRSM, the two key parameters $T_c$ and $v_c$ can be obtained from the effective potential (\ref{eqn:Veff}) by requiring the two conditions $V_{eff}(T_c; v_c)=V_{eff}(T_c; 0)$ and $v_c \neq 0$. In the numerical evaluations,  we change the temperature from a sufficiently highly energy scale, say $v_R$, toward lower values around the EW scale. A reasonable critical temperature $T_c$ for the phase transition $SU(2)_R \times U(1)_{B-L} \to \times U(1)_Y$ is assumed to be within this range. The dependence of $v_c/T_c$ on the parameters in the LRSM is exemplified in Fig.~\ref{figvT}, where in the numerical calculations  we have included all the contributions in Eq.~(\ref{eqn:Veff}).



Taking into account all the theoretical and experimental constraints in Section~\ref{sec:lrsm}, we first consider scenarios with the simplifications $\lambda_2=\lambda_3=\lambda_4=\alpha_1=\alpha_2=0$. In order to identify the parameter space where the phase transition is of first-order, we calculate $v_c/T_c$ at the critical temperature $T_c$ with different values of the quartic couplings $\rho_1$, $\rho_2$, $\rho_3 - 2 \rho_1$, $\alpha_3$. When we calculate the dependence of $v_c/T_c$ on two of the quartic couplings, all others are fixed in the way that their corresponding scalar masses equal the $W_R$ mass, and the gauge coupling $g_R = g_L$. To be concrete, we have set the renormalization scale $\mu$ to be the $v_R$ scale in Eq.~(\ref{eqn:Veff}). The corresponding results are shown in the first three panels of Fig.~\ref{figvT}. The dependence of $v_c/T_c$ on the couplings $\rho_1$ and $\alpha_3$, $\rho_3 - 2 \rho_1$ and $\rho_1$, and $\rho_2$ and $\rho_1$ are shown respectively in the upper left, upper right and lower left panels. The quantity $v_c/T_c$ is a dimensionless parameter and it is independent of the right-handed scale $v_{R}$  in the limit of $v_R \gg v_{\rm EW}$. As the quartic couplings $\rho_1$, $\rho_2$, $\rho_3 - 2 \rho_1$, $\alpha_3$ are related directly to the scalar masses $M_{H_3^0}$, $M_{H_2^{\pm\pm}}$, $M_{H_2^0}$ and $M_{H_1^0}$ (cf. Table~\ref{tab:mass}), the dependence of $v_c/T_c$ on the quartic couplings in Fig.~\ref{figvT} can also be understood effectively as the dependence of $v_c/T_c$ on the mass$-v_R$ ratios $M_{H_1^0}/v_R$, $M_{H_2^0}/v_R$, $M_{H_3^0}/v_R$ and $M_{H_2^{\pm\pm}}/v_R$. Through the gauge boson masses $M_{W_R}$ and $M_{Z_R}$, the parameter $v_c/T_c$ depends also on the gauge coupling ratio $r_g$, or equivalently on the right-handed gauge coupling $g_R$. This is shown in  the lower right panel in Fig.~\ref{figvT}; as seen in this figure, the $v_c/T_c$ limit on $\rho_1$ has a moderate or weak dependence on $r_g$, depending on the value of $\rho_1$.



Given the information on $v_c/T_c$ in Fig.~\ref{figvT},
a few more comments are now in order:
\begin{itemize}
\item As seen in Fig.~\ref{figvT}, a strong FOPT in the LRSM require a relatively small quartic coupling $\rho_1 \lesssim 0.07$ for the parameter space we are considering, which is qualitatively similar to the SM case where a light Higgs boson (say $M_h < 80 $ GeV) is needed in order to have a first-order EW phase transition~\cite{Cline:2006ts}.  It turns out that a small $\rho_1$ (and resultantly light $H_3^0$) is not only crucial for the prospects of GWs in future experiments (cf. Fig.~\ref{fig:GWcurves}), but also triggers rich phenomenology for the searches of LLPs at the high-energy colliders and dedicated detectors~\cite{Dev:2016vle, Dev:2017dui}.

\item The phase transition at the $v_R$ scale occurs when the neutral component $\Delta_R^0$ of the right-handed triplet $\Delta_R$ develops a non-vanishing VEV $v_R$. As a result, the strong FOPT  is more sensitive to the mass of $H_3^0$, or equivalently to the value of $\rho_1$, than other heavy scalar masses.
This is also clearly demonstrated in the plots of Fig.~\ref{figvT}. As seen in the upper left, upper right and lower left panels, the quartic coupling $\alpha_3$, $\rho_3 - 2\rho_1$ and $\alpha_3$ can reach up to order one, while $\rho_1 \lesssim 0.1$ in the Fig.~\ref{figvT}.



\item Although the quartic couplings $\alpha_3$, $\rho_3$ and $\rho_3 - 2\rho_1$ is less constrained by the FOPT than the critical coupling $\rho_1$,  as seen in the first three panels of Fig.~\ref{figvT}, if either of these couplings is sufficiently large, it will invalidate the strong FOPT at the $v_R$ scale, no matter how small $\rho_1$ is. Meanwhile, the white areas in the plots of Fig.~\ref{figvT} indicate that in these regions the perturbation method starts to break down and theoretical predictions become more difficult.
\end{itemize}

In Fig.~\ref{figvT} we have fixed some parameter in the LRSM and vary two of them. To see more details of the correlation of $v_c/T_c$ and the parameters in the LRSM, we take a more thorough scan of the parameter space of the LRSM. To be specific, we adopt the following ranges:
\begin{eqnarray}
	& \xi =10^{-3}, \quad
	\alpha_2=\beta_i=\lambda_{2,3,4}=0,\quad
	r_g=1,\quad
	v_R = 10\, {\rm TeV},\;\; 20 \, {\rm TeV} \,, \nonumber \\
	&\rho_1 \in [0, 0.5], \quad
	\alpha_3 \in [0, 10], \quad
	\rho_3-2\rho_1,\rho_2,y_{N} \in [0,2],\quad
	\lambda_1\in [0.13,2]
	\label{scan}
\end{eqnarray}
and apply all the theoretical and experimental constraints in Section~\ref{sec:lrsm}. Here follows some comments:
\begin{itemize}
	\item
	We have chosen $\xi = \kappa_2/\kappa_1=0.001$ in order to satisfy the theoretical constraint in Eq.~(\ref{eqn:xi}).
	\item We have chosen $\alpha_2=0$ in order to meet the requirement of the correct vacuum conditions given in Eq.~(\ref{eqn:correctvacuum}).
	\item  It is known from Fig.~\ref{figvT} that the strongly FOPT need a small $\rho_1$, therefore we have chosen $\rho_{1} < 0.5$.
	\item $\rho_3-2\rho_1$ has set to be larger than zero, as it corresponds to the masses of the left-handed triplet scalars (see Table~\ref{tab:mass}).
	\item The quartic coupling $\alpha_1$ is not a free parameter here, as it is related to $\lambda_1$ and the SM coupling $\lambda$ via Eq.~(\ref{eqn:lambda1}). As $\alpha^2/4\rho_1$ is always positive, it turn out that the quartic coupling $\lambda_1 \geq \lambda \simeq 0.13$.
	\item We have chosen two benchmark values of $10$ TeV and $20$ TeV for the right-handed scale $v_R$ to examine the dependence of FOPT on $v_R$. It turns out that the phase transition is almost independent of the values of $v_R$, as expected.
\end{itemize}

\begin{figure}[!t]
	\centering	
	\subfigure{\includegraphics[width=0.49\linewidth]{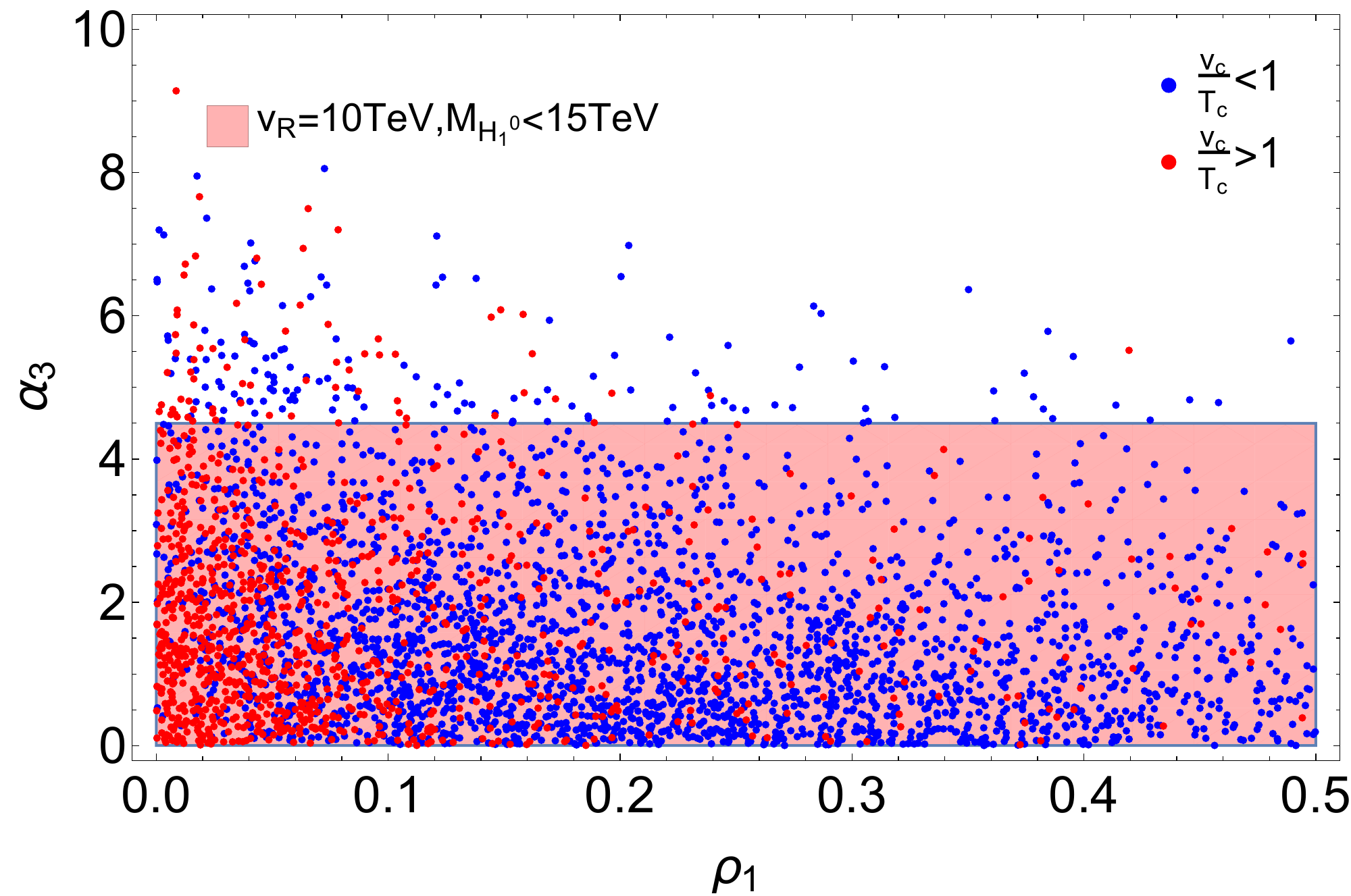}}
	\subfigure{\includegraphics[width=0.49\linewidth]{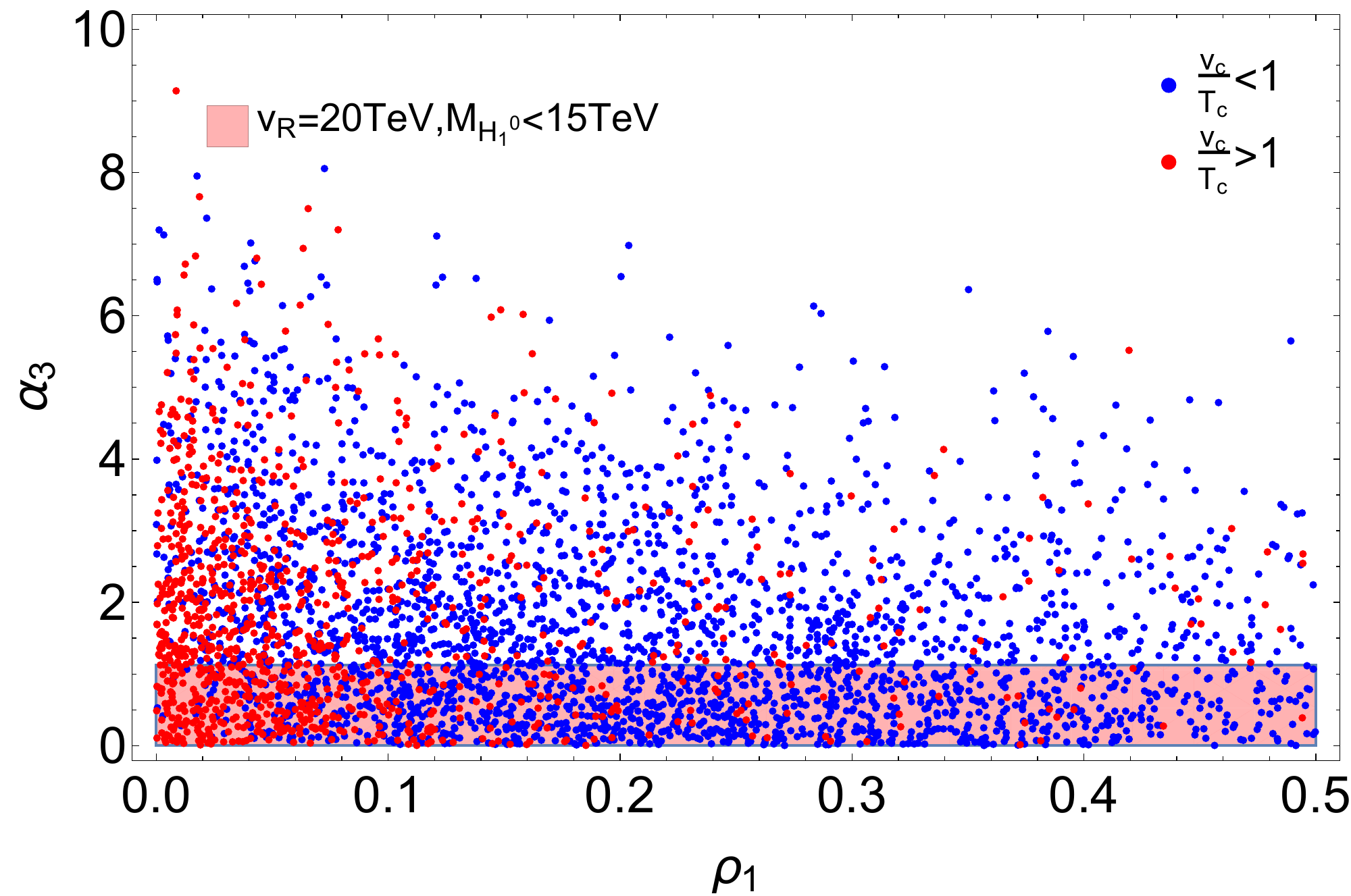}}
	\caption{Scatter plots of $\rho_1$ and $\alpha_3$, with the blue points have $v_c/T_c<1$ and the red ones $v_c/T_c>1$. In the left panel, the FCNC limits on $\alpha_3$ for $v_R = 10$ TeV and $M_{H_1^0} < 15$ TeV are indicated by the pink shaded regions. In the right panel, the case with $v_R = 20$ TeV is shown.\label{fig:random1}}
\end{figure}

The resultant scatter plots of $v_c/T_c$ are presented in Fig.~\ref{fig:random1} as functions of the parameters $\rho_1$ and $\alpha_3$. The data points of strong FOPT with $v_c/T_c>1$ are shown in red while those with $v_c/T_c<1$ are in blue. When we set $v_R = 10$ TeV and take the FCNC limit of $M_{H_1^0} > 15$ TeV~\cite{Zhang:2007da}, the quartic coupling $\alpha_3$ should meet the condition $\alpha_3 > 2 M_{H_1^0}^2/v_R^2 = 4.5$. The region shaded by the light pink in the left panel of Fig.~\ref{fig:random1} is excluded by such conditions. It is found that only a small amount of the data points can survive and have strong FOPT. When the $v_R$ scale is higher, say $v_R = 20$ TeV, the quartic coupling $\alpha_3$ is significantly smaller, i.e. $\alpha_3 > 1.13$. The region denoted by the light pink shaded region in the right panel of Fig.~\ref{fig:random1} is excluded. Then there will be more points that can have a strong FOPT with  $v_c/T_c>1$, as clearly shown in the right panel of Fig.~\ref{fig:random1}.



\section{Gravitational waves}
\label{sec:GW}

The thermal stochastic GWs can be generated by three physics processes in phase transition~\cite{Caprini:2015zlo}:  collisions of bubbles, sound waves (SWs) in the plasma after the bubble collision, and the MHD turbulence  forming after the bubble collision.  For non-runaway scenarios, GWs are dominated by the latter two sources~\cite{Caprini:2015zlo}, and the corresponding GW spectrum can be approximated as
\begin{equation}
	h^2\Omega_{\rm GW} \ \simeq \ h^2\Omega_{\rm SW}+h^2\Omega_{\rm MHD} \,.
\end{equation}

The SW contribution has the form of~\cite{Hindmarsh:2015qta}
\begin{eqnarray}
	h^2\Omega_{\rm SW} (f) & \ \simeq \ & 2.65 \times 10^{-6}\left(\frac{H_*}{\beta}\right)\left(\frac{\kappa_v \alpha}{1+\alpha}\right)^2\left(\frac{100}{g_*}\right)^{1/3} v_w \left( \frac{f}{f_{\rm SW}}\right) ^3\left[ \frac{7}{4+3 \left(\frac{f}{f_{\rm SW}}\right)^2} \right]^{7/2} \,, \nonumber \\ &&
\end{eqnarray}
where $f$ is the frequency, $g_\ast$ and $H_\ast$ are respectively the number of relativistic degrees of freedom in the plasma and the Hubble parameter at  the temperature $T_\ast$, $v_w$ is the bubble wall velocity, $\alpha$ describes the strength of phase transition, $\beta/H_{*}$ measures the rate of the phase transition, and
\begin{equation}
	\kappa_v=\frac{\alpha}{0.73+0.083\sqrt{\alpha}+\alpha},
\end{equation}
is the fraction of vacuum energy that is converted to bulk motion. The peak frequency $f_{\rm SW}$ is approximated by
\begin{eqnarray}
	f_{\rm SW} & \ \simeq \ & 1.9\times 10^{-2}\frac{1}{v_w}\left(\frac{\beta}{H_*} \right)\left(\frac{T_*}{100\text{GeV}} \right)   \left(\frac{g_*}{100} \right)^{1/6} \text{mHz} \,.
\end{eqnarray}

The MHD turbulence contribution is~\cite{Caprini:2009yp, Binetruy:2012ze}
\begin{eqnarray}
	h^2\Omega_{\rm MHD} (f) & \ \simeq \ & 3.35 \times 10^{-4}\left(\frac{H_*}{\beta}\right)\left(\frac{\kappa_{\rm MHD} \alpha}{1+\alpha}\right)^{3/2} \left(\frac{100}{g_*}\right)^{1/3} v_w \frac{\left( \frac{f}{f_{\rm MHD}} \right)^3}{\left( 1+ \frac{f}{f_{\rm MHD}}\right)^{11/3} \left( 1+ \frac{8\pi f}{h_*} \right)} \,, \nonumber \\ &&
	\label{omigaturb}
\end{eqnarray}
where $\kappa_{\rm MHD} \simeq 0.05\kappa_v$ is the fraction of vacuum energy that is transformed into the MHD turbulence, $h_\ast$ is the inverse Hubble time at the GW production (red-shifted to today), and is given by
\begin{equation}
	h_*=16.5\times 10^{-6}\left(\frac{T_*}{100\text{GeV}} \right)   \left(\frac{g_*}{100} \right)^{1/6} \text{Hz} \,,
\end{equation}
and the peak frequency is
\begin{eqnarray}
	f_{\rm MHD} & \ \simeq \ & 2.7\times 10^{-2}\frac{1}{v_w}\left(\frac{\beta}{H_*} \right)\left(\frac{T_*}{100\text{GeV}} \right)   \left(\frac{g_*}{100} \right)^{1/6} \text{mHz} \,.
\end{eqnarray}


As shown in the formula above, the gravitational wave spectrum from FOPTs are generally characterized by two parameters related to the phase transition, namely $\alpha$ and $\beta$ \cite{Grojean:2006bp}. The parameter $\alpha$ is defined as the ratio of the vacuum energy density $\epsilon_\ast$ released at the phase transition temperature $T_*$ to the energy density of the universe in the radiation era, i.e.
\begin{eqnarray}
\alpha \ = \ \frac{\epsilon_{*}}{g_{*}\pi^2 T_{*}^4 / 30} \,,
\end{eqnarray}
where $\epsilon_{*}$ is the latent heat and can be expressed as
\begin{eqnarray}
	\epsilon_{*} \ = \ \left.\left( - \Delta V_{\rm eff} +T \frac{d \Delta V_{\rm eff}}{d T} \right) \right|_{T=T_{*}} \,.
\end{eqnarray}
The $\Delta V_{\rm eff}$ denotes the difference of potential energy between the false vacuum and true vacuum, i.e. $\Delta V_{\rm eff} = - V_{\rm eff} (0, T) + V_{\rm eff} (v, T)$, which can be simply determined by $T_*$ and the parameters of LRSM.

The parameter $\beta$ describes the rate of variation of the bubble nucleation rate during phase transition, and its inverse describes the duration of phase transition. To describe rate of the phase transition, a dimensionless parameter $\frac{\beta}{H^*}$ is defined from the following equation
  \begin{equation}
  	\frac{\beta}{H_{*}}=\left. T \frac{d(S_3/T)}{dT}\right|_{T=T_{*}},
  \end{equation}
where $S_3$ denotes the three-dimensional Euclidean action of a critical bubble. The $T_*$ denotes the temperature when the phase transition is ended and can be determined by requiring that the probability for nucleating one bubble per horizon volume equals 1, i.e.
\begin{equation}
   \int_{T_*}^{T_c}\frac{dT}{T}\frac{\Gamma(T)}{H^4}=1 \,,
\end{equation}
where $\Gamma(T)$ is the probability of bubble nucleation per horizon volume, which can be expressed as $\Gamma(T) = \Gamma_0 \exp\{- {S_3}/{T} \}$, with $\Gamma_0 = T^4 (S_3/2\pi T)^{3/2}$~\cite{Coleman:1977py,Linde:1980tt,Linde:1981zj}. In this paper, $S_3$ is computed using the code {\tt CosmoTransitions}~\cite{Wainwright:2011kj} to solve the bounce equation of bubbles.

\begin{figure}[!t]
 \includegraphics[width=0.49\linewidth]{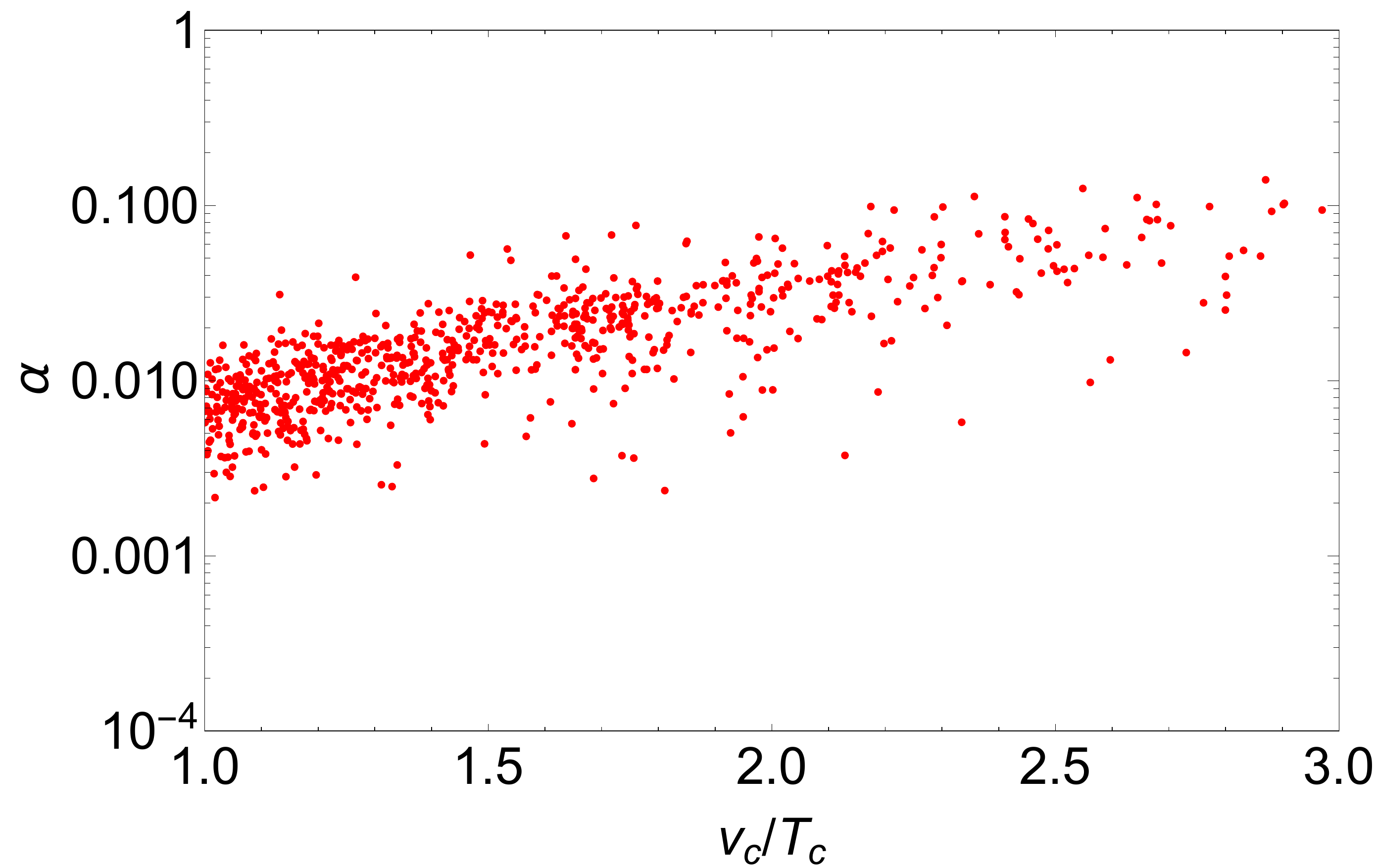}
 \includegraphics[width=0.5\linewidth]{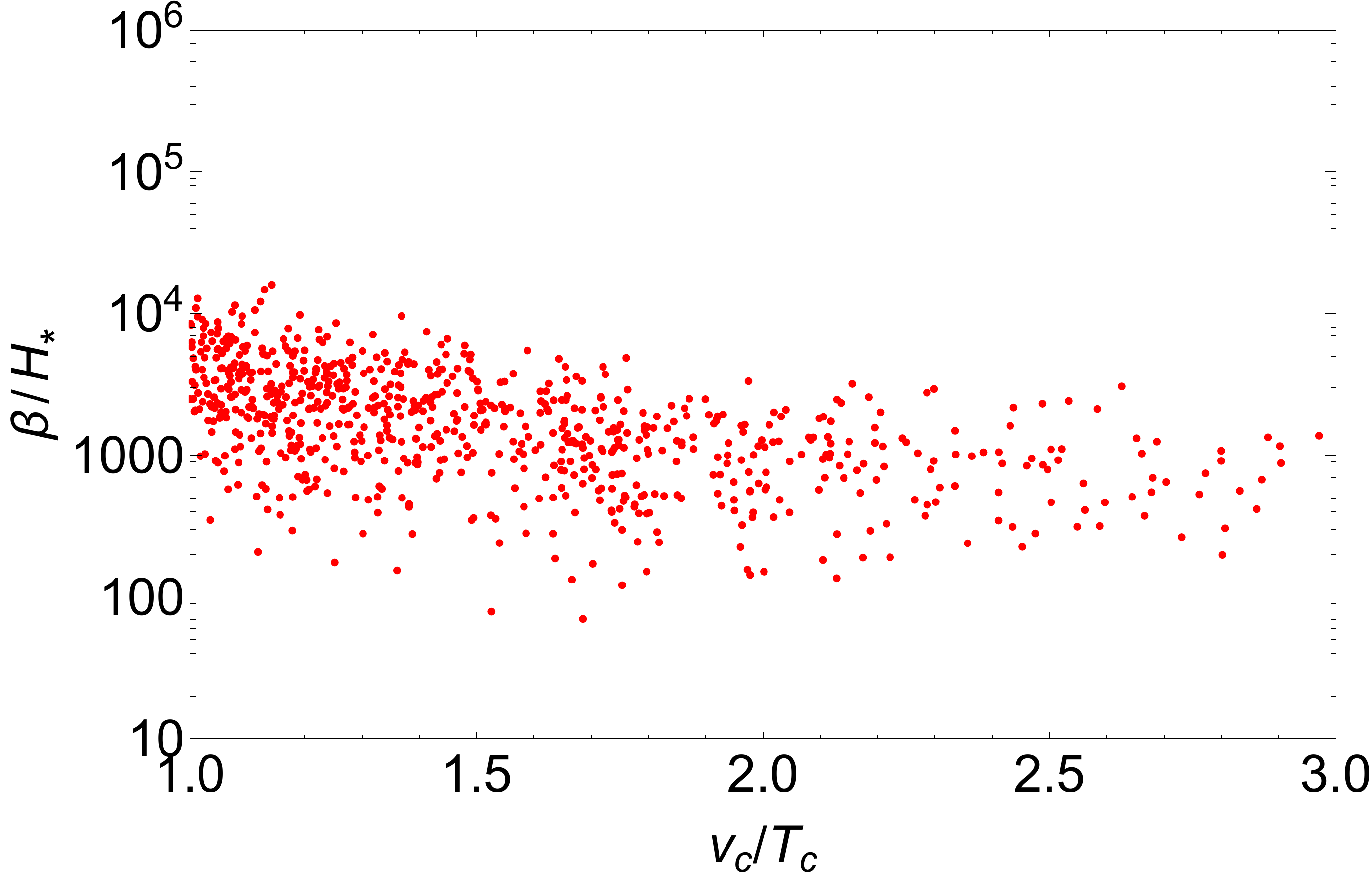}
 \caption{The values of $\alpha$ (left) and $\beta/H_*$ (right) for data points  which have strong FOPT, as function of $v_c/T_c$.}
 \label{alphabetavrTc}
\end{figure}

The parameters $\alpha$ and $\beta$ set respectively the strength and time variation of GWs during the phase transition, and their typical values in the LRSM are shown  respectively in the left and right panels of Fig.~\ref{alphabetavrTc}. As demonstrated by the data points, the value of $\alpha$ varies roughly from $0.001$ to $0.1$, and  $\beta / H_\ast$ can range from $10^2$ to $10^4$. In the numerical calculations, all the data points in Fig.~\ref{alphabetavrTc} have strong FOPT.

\begin{figure}[!t]
		\includegraphics[height=0.3\linewidth]{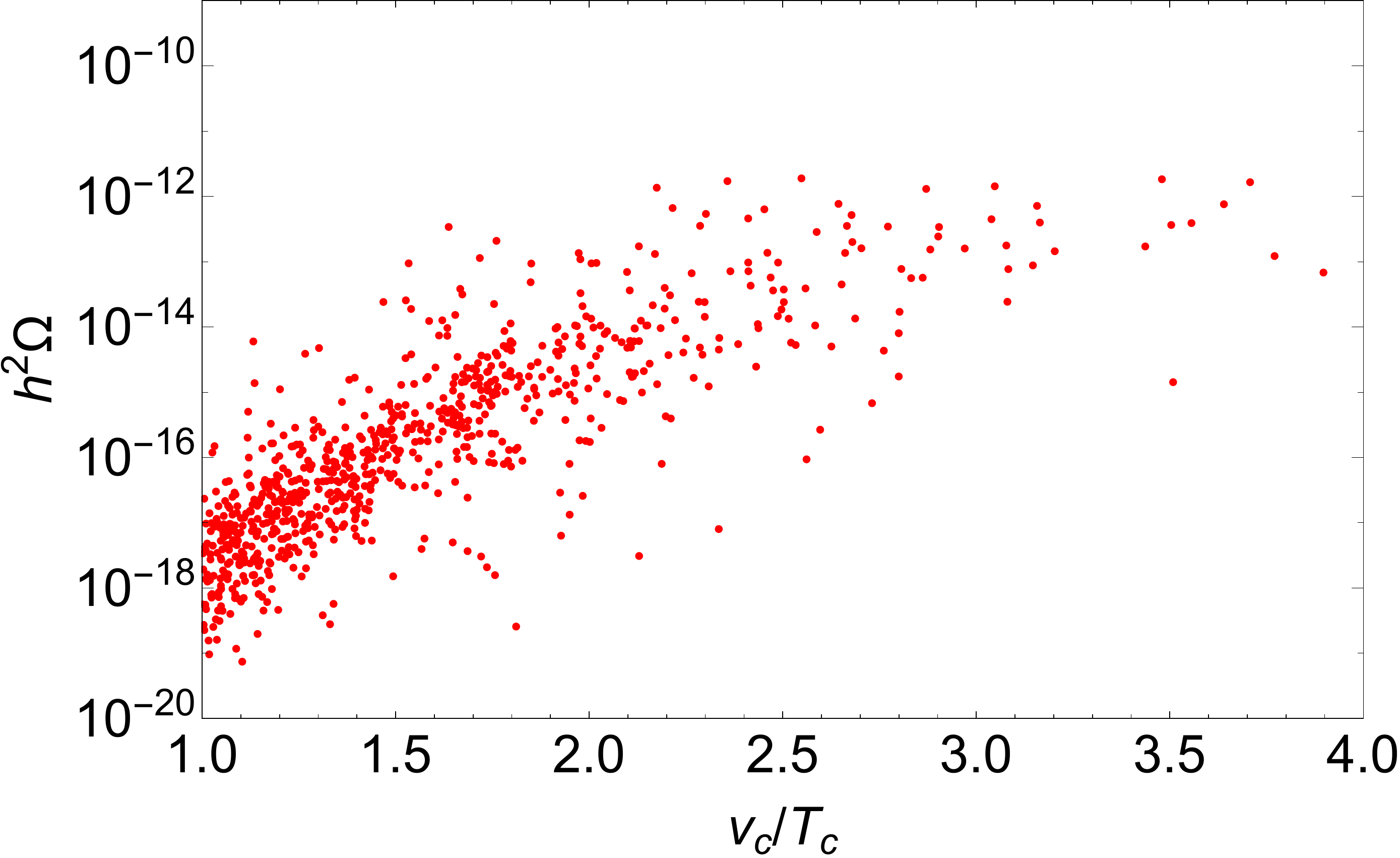}
		\includegraphics[height=0.305\linewidth]{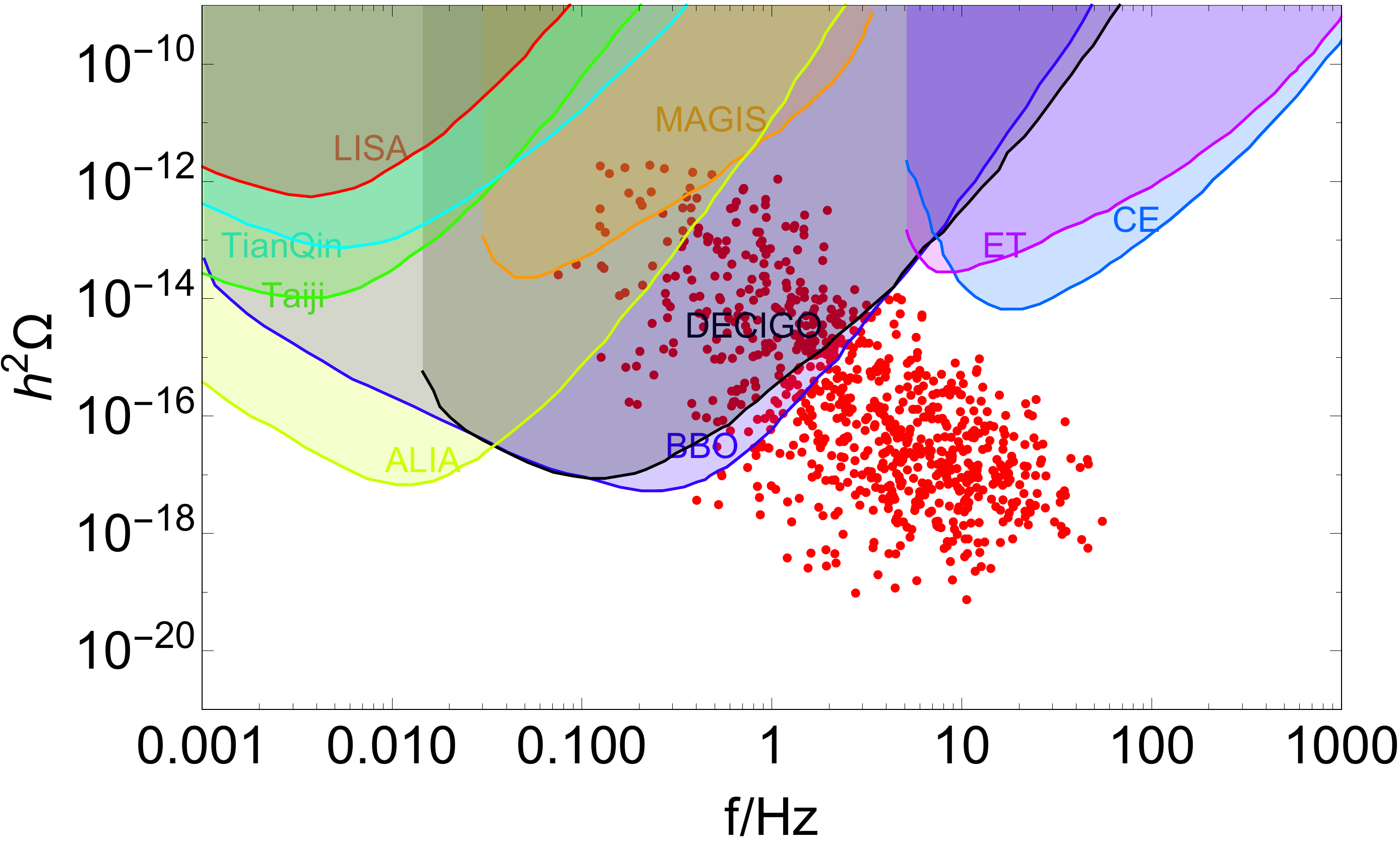}
	\caption{  GW peaks for the data points in Fig.~\ref{alphabetavrTc}, as function of $v_c/T_c$ (left) and frequency $f$ (right). Also shown in the right panel are the prospects of LISA~\cite{Audley:2017drz, Cornish:2018dyw}, TianQin~\cite{Luo:2015ght}, Taiji~\cite{Guo:2018npi},  ALIA~\cite{Gong:2014mca}, MAGIS~\cite{Coleman:2018ozp}, BBO~\cite{Corbin:2005ny},  DECIGO~\cite{Musha:2017usi}, ET~\cite{Punturo:2010zz}, and CE~\cite{Evans:2016mbw}.}	
	\label{GWpeak}
\end{figure}

Assuming the bubble wall velocity $v_w\sim 1$, the corresponding GW signals of the data points in Fig.~\ref{alphabetavrTc} are shown in Fig.~\ref{GWpeak}. The correlation of the ratio $v_c/T_c$ and GW signal peaks are presented in the left panel. We can read from Fig.~\ref{alphabetavrTc} and the left panel of Fig.~\ref{GWpeak} that with large $v_c/T_c$ the value $\alpha$ is typically larger, thus yielding stronger GW signals. The GW strength and frequency peaks are shown in the right panel of Fig.~\ref{GWpeak}. The potential sensitivities of LISA~\cite{Audley:2017drz, Cornish:2018dyw}, TianQin~\cite{Luo:2015ght}, Taiji~\cite{Guo:2018npi},  ALIA~\cite{Gong:2014mca}, MAGIS~\cite{Coleman:2018ozp}, BBO~\cite{Corbin:2005ny},  DECIGO~\cite{Musha:2017usi}, ET~\cite{Punturo:2010zz}, and CE~\cite{Evans:2016mbw} are also depicted in the right panel of Fig.~\ref{GWpeak}. As seen in this figure, the frequency peak in the LRSM can range from $10^{-1}$ to $10^2$ Hz.
Furthermore, there are some data points of the LRSM with frequencies in the range of roughly from $0.1$ to $10$ Hz and GW strength larger than $10^{-17}$,  which can be detected in the future by BBO and DECIGO, or even by ALIA and MAGIS.

\begin{figure}[!t]
	\centering
	\subfigure{\includegraphics[width=0.49\linewidth]{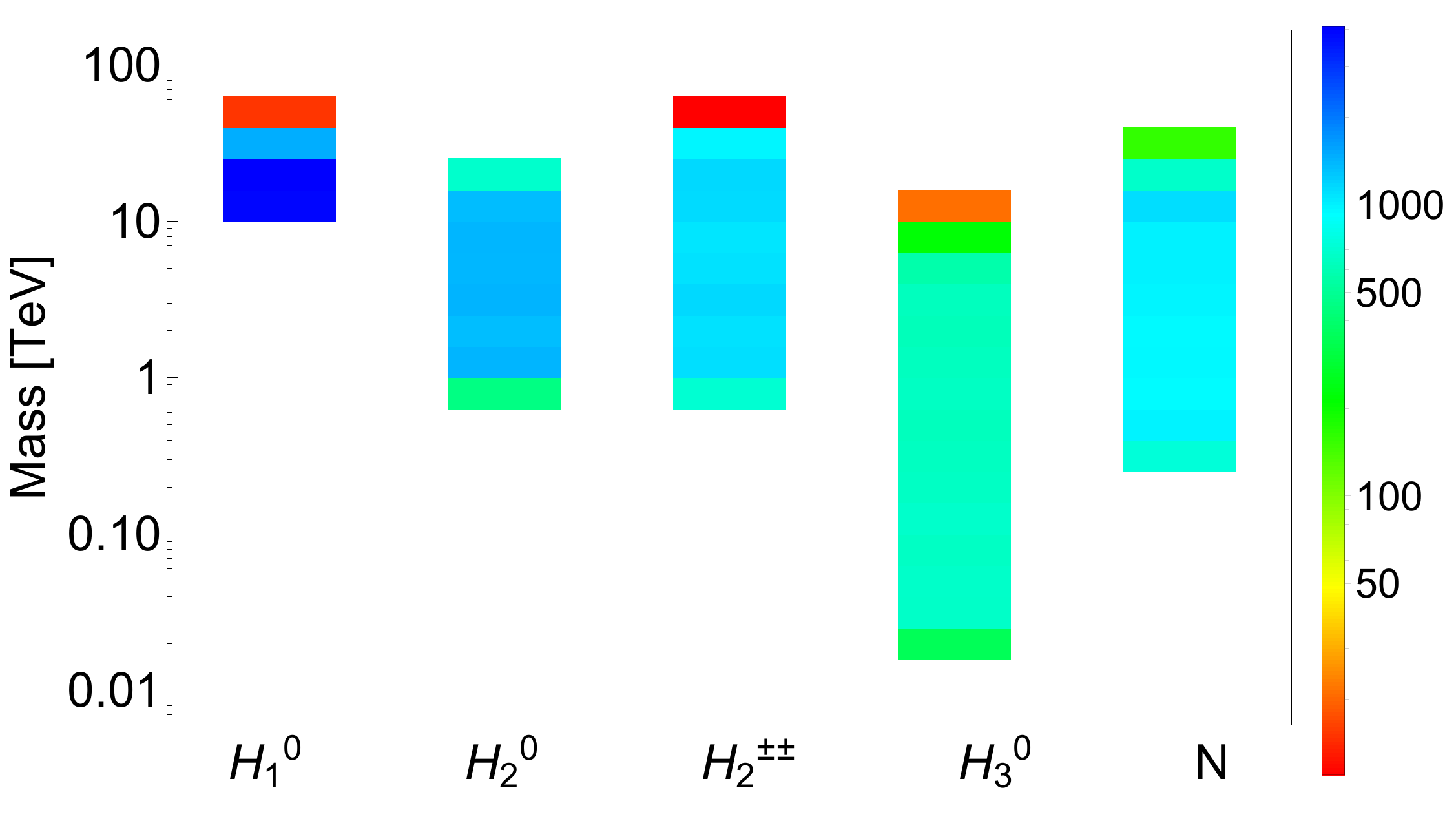}}
	\subfigure{\includegraphics[width=0.49\linewidth]{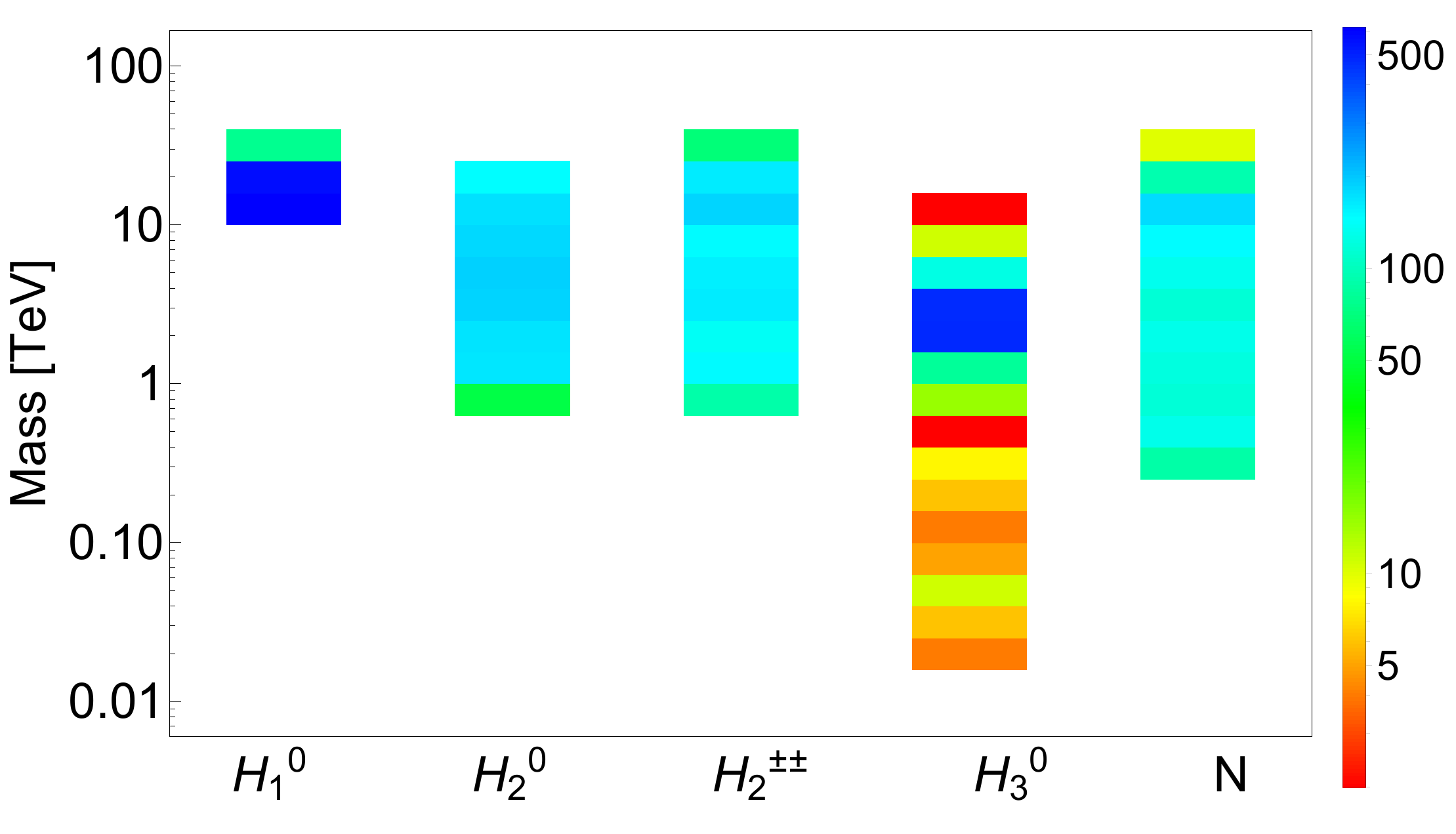}}
	\caption{ Distributions of data points as function of the masses of $H_1^0$, $H_2^0$, $H_2^{\pm\pm}$, $H_3^0$ and $N$, with the strong FOPT  $v_c/T_c>1$ (left), and for the data points that can be detected by BBO and DECIGO (right).}
	\label{fig:massofsamples}
\end{figure}	

\begin{figure}[!t]
	\centering
	\includegraphics[width=0.6\linewidth]{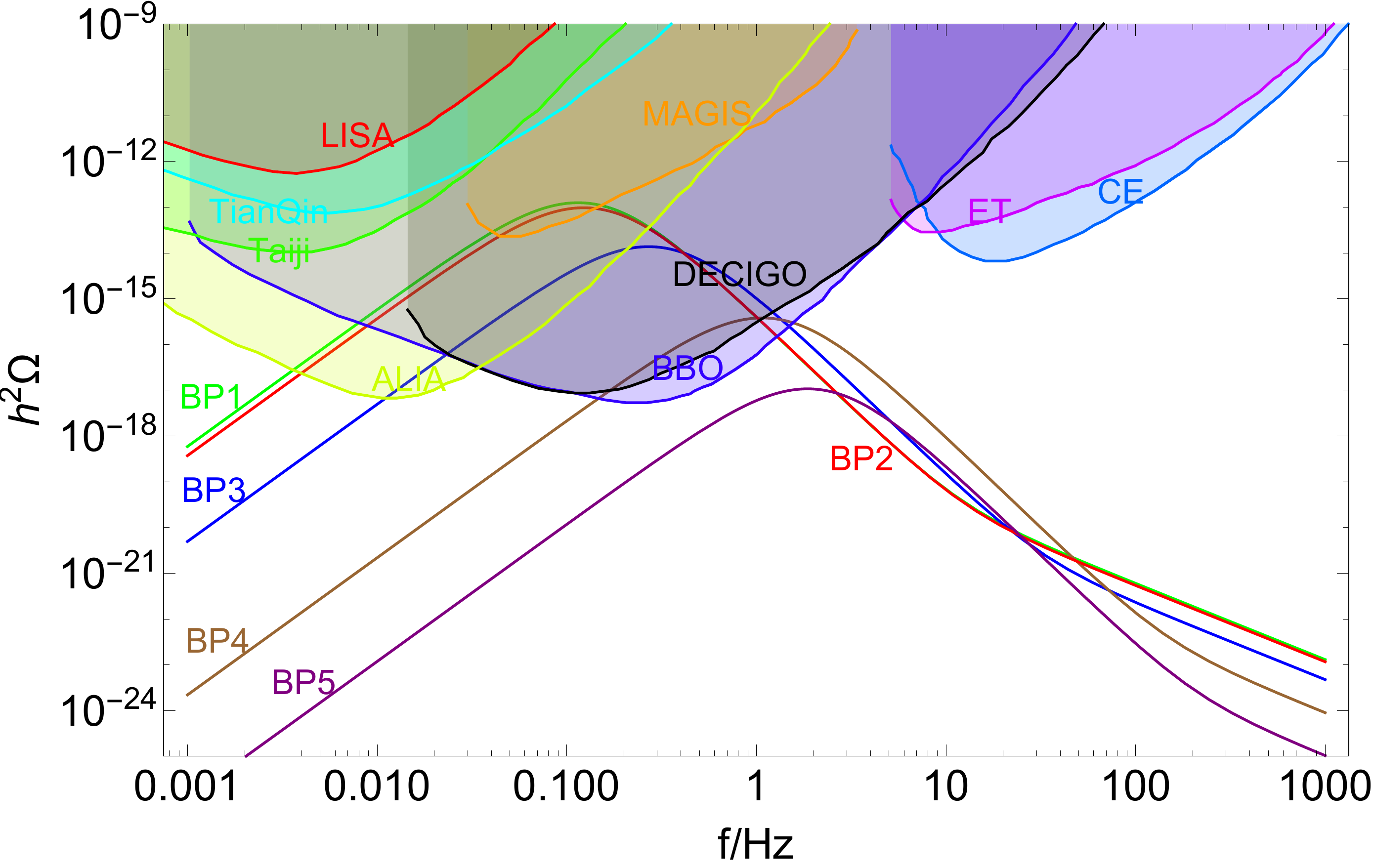}
	\caption{The same as in the right panel of Fig.~\ref{GWpeak}, but for the five BPs in Table~\ref{tab:BPs}.}
	\label{fig:GWcurves}
\end{figure}

For the data points in Fig.~\ref{GWpeak} with strong FOPT, the mass spectra of the scalars $H_1^0$, $H_2^0$, $H_3^0$, $H_2^{\pm\pm}$ and the mass of RHNs $N$ are shown in the left panel of Fig.~\ref{fig:massofsamples}, and the mass spectra of these particles for the data points that are achievable in the BBO and DECIGO experiments are presented in the right panel of Fig.~\ref{fig:massofsamples}. The two plots of Fig.~\ref{fig:massofsamples} clearly show that the masses of $H_1^0$, $H_2^0$ and $H_2^{\pm\pm}$ can reach up to few times 10 TeV, with their lower mass limits roughly round the experimental constraints in Section~\ref{sec:experimental} (see also Table~\ref{table:bounds} and Fig.~\ref{fig:spectra}). The mass of $H_3^0$ can go to much smaller values, i.e. from 20 GeV up to 10 TeV. This can be easily understood: on one hand, the theoretical and experimental constraints on $H_3^0$ mass are rather weak (see Section~\ref{sec:lrsm}); on the other hand, the strong FOPT and GW production in the LRSM favor a relatively light $H_3^0$ (see Figs.~\ref{figvT}, \ref{fig:random1} and \ref{fig:GWcurves}).
As seen in Fig.~\ref{fig:massofsamples}, the RHN masses $M_N$ can range roughly from 300 GeV up to 40 TeV. It is expected that the GW probe of $H_3^0$ and RHNs are largely complementary to the direct searches of them at the high-energy colliders, including the searches of long-lived $H_3^0$ and $N$. See Section~\ref{sec:H3} for more details.






\begin{figure}[!t]
	\centering
	\includegraphics[width=0.75\textwidth]{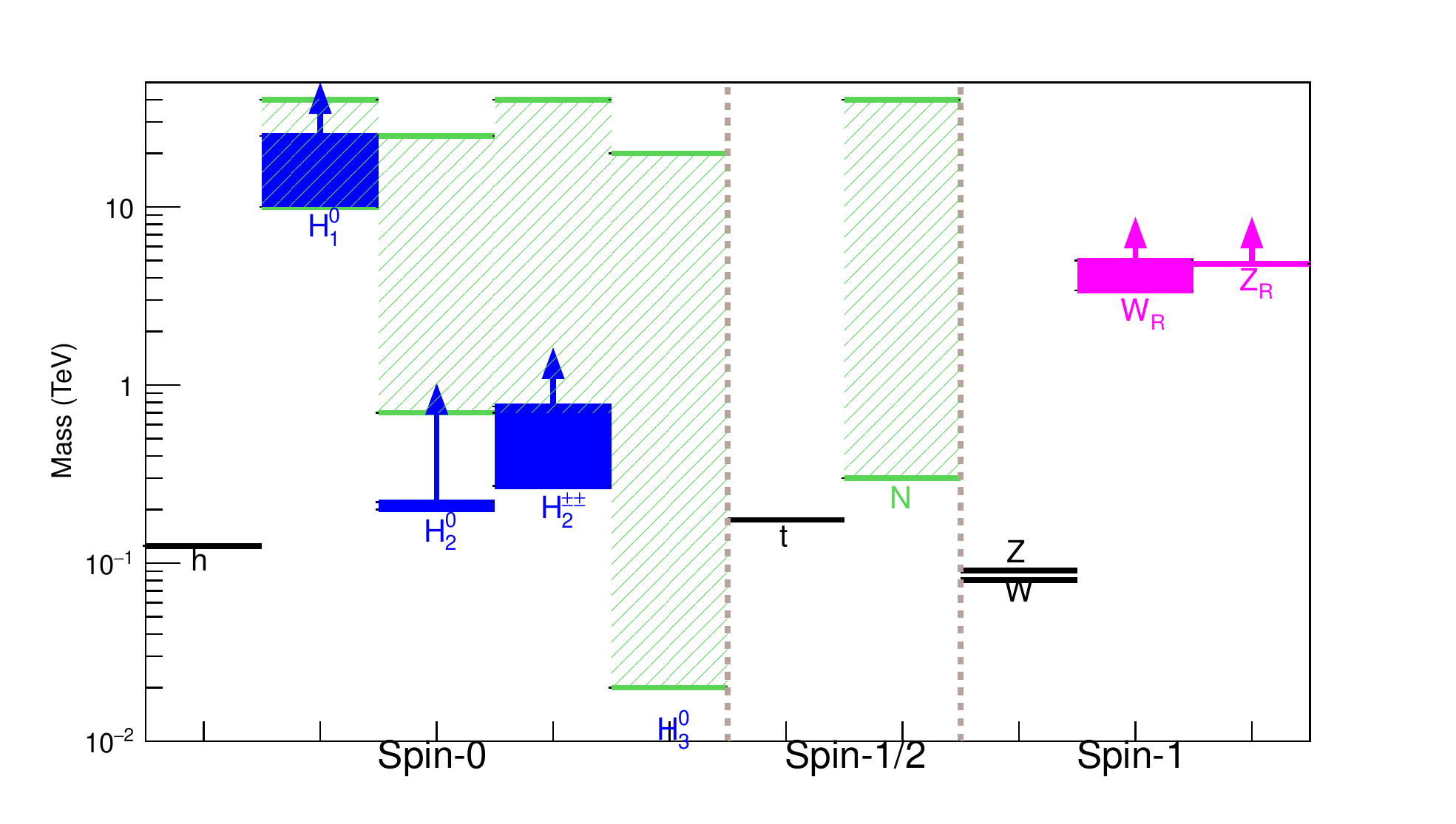}
	\caption{Combined plot of the experimental limits in Fig.~\ref{fig:spectra} (blue and pink blocks with arrows) and the GW prospects of the masses of $H_1^0$, $H_2^0$, $H_2^{\pm\pm}$, $H_3^0$ and $N$ in the right panel of Fig.~\ref{fig:massofsamples}  (green hatched regions). The horizontal black lines are the masses of SM Higgs $h$, top quark $t$, and $W$, $Z$ bosons.
	\label{fig:cmbspectra} }
\end{figure}

For the purpose of comparison, we present in Fig.~\ref{fig:cmbspectra} the experimental limits on the masses of $H_1^0$, $H_2^0$ and $H_2^{\pm\pm}$ in Fig.~\ref{fig:spectra} and the GW sensitive ranges of the masses of $H_1^0$, $H_2^0$, $H_2^{\pm\pm}$, $H_3^0$ and $N$ in Fig.~\ref{fig:massofsamples}, where the mass ranges within the sensitivities of GW detectors are represented by green hatched areas. It is clear that the GWs from phase transition can probe a large region of parameter space in the LRSM that goes beyond the current collider limits.

To expose more features of GWs from the phase transition at the $v_R$ scale in the LRSM, we have chosen five specific BPs. For the sake of concreteness and simplification, we have chosen $v_R =10$ TeV, $\xi = 10^{-3}$, and set the quartic couplings $\lambda_1= \lambda = 0.13$, $\alpha_1=\alpha_2=\lambda_2=\lambda_3=\lambda_4=0$. The BSM particle masses $M_{H_1^0}$, $M_{H_2^0}$, $M_{H_3^0}$, $M_{H_2^{\pm\pm}}$ and $M_N$ are collected in the first few columns of Table~\ref{tab:BPs}. The resultant $v_c$, $T_c$, $T_\ast$ and the parameters $\alpha$ and $\beta/H_\ast$ are also shown in Table~\ref{tab:BPs}. The GW spectra $h^2 \Omega$ as function of the frequency $f$ for the five BPs are presented in Fig.~\ref{fig:GWcurves}.
There are a few comments on the five BPs.

\begin{table}[!t]
	\begin{center}
		\caption{Five BPs studied in this paper. Parameters not shown in the table are set to be $v_R=10$ TeV, $\xi=10^{-3}$, $\lambda_1=0.13$, $\alpha_1=\alpha_2=\lambda_2=\lambda_3=\lambda_4=0$. Their GW spectra are shown in Fig.~\ref{fig:GWcurves}. It is also noticed that all these BPs are non-runaway scenarios in term of the criteria defined in Eq.~(25) of \cite{Caprini:2015zlo}. The suppression factor $\Upsilon$ in the last row is defined in Eq.~(\ref{eqn:upsilon})~\cite{Guo:2020grp,Fornal:2020esl}. }
		\label{tab:BPs}
		\vspace{5pt}
		\begin{tabular}{cccccc}
			\hline\hline
			BPs & BP1 & BP2 & BP3 & BP4 & BP5 \\ \hline
			$M_{H_1^0}$ & 10 TeV & 10 TeV & 10 TeV & 10 TeV & 10 TeV \\ \hline
			$M_{H_2^0}$ & 8 TeV & 8 TeV & 8 TeV & 8 TeV & 10 TeV \\ \hline
			$M_{H_3^0}$ & 40 GeV & 500 GeV & 1 TeV & 2 TeV & 2 TeV \\ \hline
			$M_{H_2^{\pm\pm}}$ & 8 TeV & 8 TeV & 8 TeV & 8 TeV & 10 TeV \\ \hline
			$M_{N}$ & 1 TeV & 1 TeV & 1 TeV & 1 TeV & 2 TeV \\ \hline
			$v_c$ & 8.02 TeV & 8.01 TeV & 7.98 TeV & 7.72 TeV & 7.18 TeV  \\ \hline
			$T_c$ & 3.42 TeV & 3.50 TeV & 3.73 TeV & 4.49 TeV & 5.44 TeV \\ \hline
			$T_*$ & 2.17 TeV & 2.27 TeV & 2.75 TeV & 3.92 TeV & 4.89 TeV \\ \hline
			$\alpha$ & 0.056 & 0.053 & 0.037 & 0.019 & 0.0083 \\ \hline
	                $\alpha_\infty$ & 0.18 & 0.16 & 0.11 & 0.053 & 0.037 \\ \hline
			$\beta/H_*$ & 265 & 272 & 493 & 1373 & 1908 \\ \hline
			$\Upsilon$ & 0.16 & 0.16 & 0.13 & 0.10 & 0.15\\
			\hline\hline
		\end{tabular}
	\end{center}
\end{table}

\begin{itemize}
	\item It is clear in Fig.~\ref{fig:GWcurves} that the BPs (from BP1 to BP4) with the same values of $M_{H_1^0}$, $M_{H_2^0}$, $M_{H_2^{\pm\pm}}$ and $M_N$ but different $M_{H_3^{0}}$ can be probed in the future by BBO and DECIGO, and even by ALIA and MAGIS. It seems that the $H_3^0$ mass $M_{H_3^0}$, or equivalently the quartic coupling $\rho_1$, is crucial for the GWs in the LRSM. The BPs (like BP5) with a heavier $H_3^0$, or equivalently larger $\rho_1$, tends to generate a small $\alpha$ and large $\beta$, and thus produce weaker GW signals with a larger frequency. This is consistent with the findings in Ref.~\cite{Brdar:2019fur}.
	The BPs BP1 and BP2 with $H_3^0$ mass below TeV-scale can produce GWs of order $10^{-13}$ with frequency at around 0.1 Hz, far above the prospects of BBO and DECIGO. The BP4 with a 2 TeV $H_3^0$ can only produce GWs of order $10^{-16}$ with frequency peaked at 1 Hz, which can be marginally detected by BBO and DECIGO.
	
	\item Comparing BP4 and BP5, it is clear that only the masses of $H_2^{0}$, $H_2^{\pm\pm}$ and $N$ are heavier in BP5 than in BP4, while all other parameters are the same. As seen in Fig.~\ref{fig:GWcurves}, the GW signal in BP5 is so weak that it can escape the detection of all the planned GW experiments in the figure. This reveals that the masses $M_{H_2^{0}}$, $M_{H_2^{\pm\pm}}$ and $M_N$, or equivalently the couplings $\rho_3 - 2\rho_1$, $\rho_2$ and $y_N$, are also important for GW production in the LRSM. More data points in the numerical calculations reveal that the coupling $\alpha_3$ is also very important for the GW signals in the LRSM.
	
\end{itemize}

\section{Complementarity of GW signal and collider searches of LRSM}
\label{sec:complementarity}

In spite of the large number of BSM scalars, fermions and gauge bosons in the LRSM and  the larger number of quartic couplings in the potential~(\ref{eqn:potential}), it is phenomenologically meaningful to examine the role of some couplings, or equivalently the BSM particle masses, in the strong FOPT and the subsequent GW production in the early universe, as well as the potential correlations of GWs to the direct laboratory searches of these particles and the SM precision data at the high-energy colliders. In this section, we will elaborate on (i) the effects of the quartic coupling $\lambda_1$ in the scalar potential (\ref{eqn:potential}) which corresponds to the self-coupling $\lambda$ in the SM, and (ii) the complementarity of GW signal, the collider searches of (light) $H_3^0$ and the heavy (or light) RHNs in the LRSM.

\subsection{Self-couplings of SM-like Higgs boson in the LRSM}

\begin{table}[!]
	\begin{center}
		\caption{Comparison of the masses squared, trilinear and quartic couplings of the SM-like Higgs $h$ in the SM and LRSM~\cite{Dev:2016dja,  Maiezza:2016ybz}.
		\label{hselfcpl}}
	    \vspace{5pt}
		\begin{tabular}{cccc}
			\hline\hline
			models & mass squared & $\lambda_{hhh}$ & $\lambda_{hhhh}$   \\ \hline
			SM & $2 \lambda^{} v_{\rm EW}^2$& $\lambda^{} v_{\rm EW} $ & $\frac14 \lambda^{}$\\ \hline
			LRSM& $(2 \lambda_1 - \frac{\alpha_1^2}{2\rho_1}) v_{\rm EW}^2$ & $\frac{1}{4}\left(4\lambda_1-\frac{\alpha_1^2}{\rho_1}\right)v_{\rm EW} + \left(4\lambda_4-\frac{\alpha_1\alpha_2}{\rho_1}\right)\xi v_{\rm EW}$ & $\frac14 {\lambda_{1}}$ \\
			\hline\hline
		\end{tabular}
	\end{center}
\end{table}

It is interesting to examine how the self-coupling $\lambda$ of the SM-like Higgs boson $h$ can be affected by the BSM scalars in the LRSM. The SM-like Higgs mass square, the trilinear coupling $\lambda_{hhh}$ and the quartic coupling $\lambda_{hhhh}$ in the SM and LRSM are collected in Table~\ref{hselfcpl}. Comparing the mass square of $h$ in the SM and LRSM, we can approximately identify the following relation among the SM and LRSM quartic couplings~\cite{Dev:2016dja, Maiezza:2016ybz}
\begin{eqnarray}
	\label{eqn:lambda1}
	\lambda_1 - \frac{\alpha_1^2}{4\rho_1} \simeq \lambda \,.
\end{eqnarray}
As seen in the third column of Fig.~\ref{hselfcpl}, the trilinear coupling $\lambda_{hhh}$ of the SM-like Higgs in the LRSM only differs from the SM value by a small amount of $\xi\sim10^{-3}$~\cite{Dev:2016dja, Maiezza:2016ybz}. On the contrary, the quartic coupling $\lambda_{hhhh}$ in the LRSM might be significantly different from the SM prediction: as shown in the last column of Table~\ref{hselfcpl}~\cite{Dev:2016dja,  Maiezza:2016ybz},
\begin{eqnarray}
	\label{eqn:diff}
	\frac14 \lambda_1 - \frac14 \lambda \simeq \frac{\alpha_1^2}{16\rho_1} \,.
	\label{ldev}
\end{eqnarray}
In other words, at the leading-order of the approximations of $v_R \gg v_{\rm EW} \simeq \kappa_1 \gg \kappa_2$, the difference of quartic coupling of SM-like Higgs boson in the SM and LRSM is dominated by the $\alpha_1^2/16\rho_1$ term. As the FOPT and GW in the LRSM favor a small $\rho_1$ coupling, the difference in Eq.~(\ref{eqn:diff}) tends to be significant for sufficiently large $\alpha_1$.

\begin{figure}[!t]
	\centering	
	\subfigure{\includegraphics[width=0.32\linewidth]{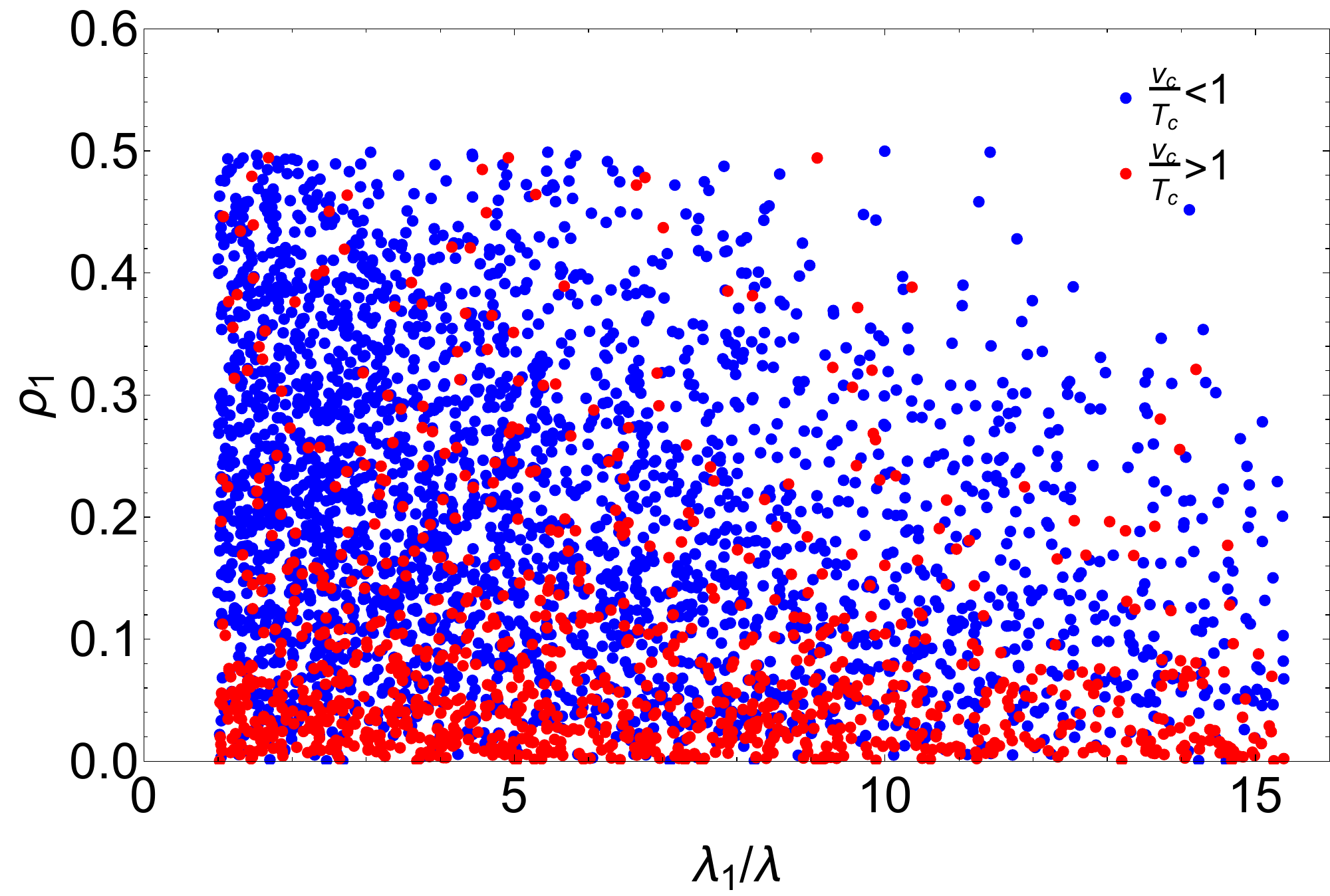}}
	\subfigure{\includegraphics[width=0.32\linewidth]{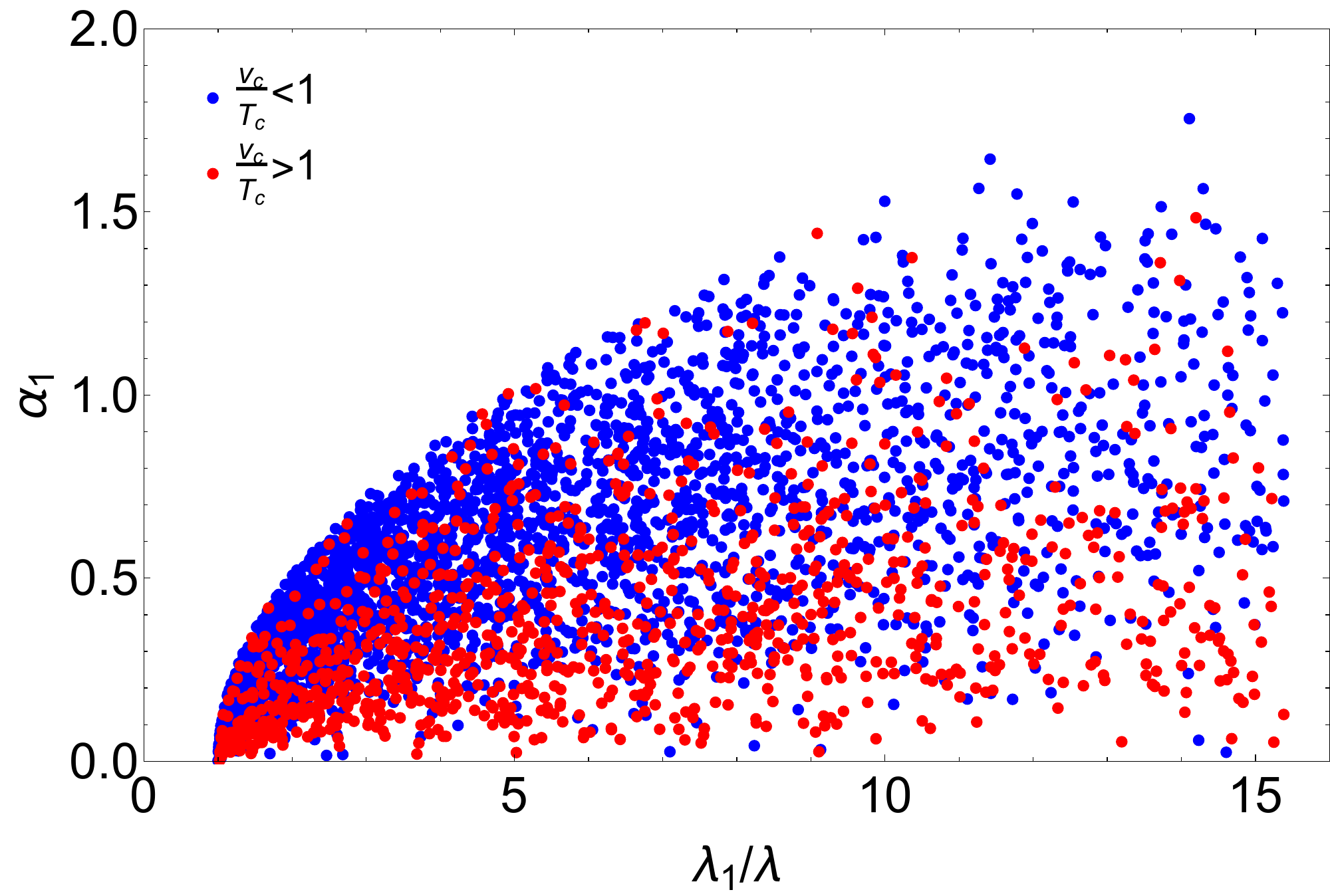}}
	\subfigure{\includegraphics[width=0.32\linewidth]{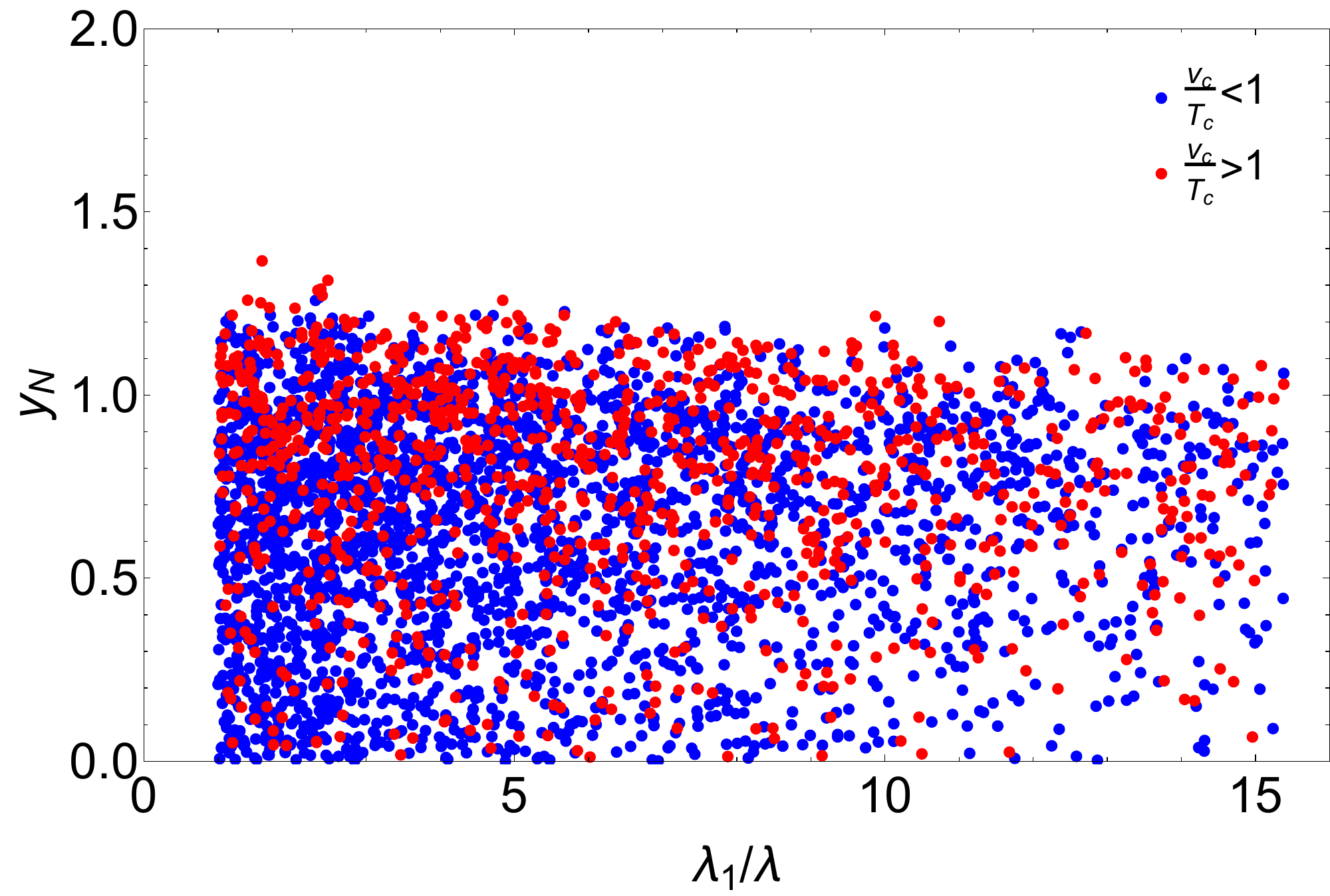}}
	\caption{Scatter plots of $\lambda_1/\lambda$ and $\rho_1$ (left), $\alpha_1$ (middle) and $y_N$ (right), with the blue points have $v_c/T_c<1$ and the red ones $v_c/T_c>1$.  
	}
	\label{fig:random2}
\end{figure}

Adopting the parameter ranges in Eq.~(\ref{scan}) and taking into account the theoretical and experimental limits in Section~\ref{sec:lrsm}, the scatter plots of the quartic coupling $\lambda_{hhhh}$ and the couplings $\rho_1$, $\alpha_1$ and $y_N$ are shown respectively in the left, middle and right panels of Fig.~\ref{fig:random2}, where the data points with strong FOPT $v_c/T_c>1$ is shown in red, while those with $v_c/T_c<1$ are in blue. It is very clear in Fig.~\ref{fig:random2} that the deviation of the quartic scalar coupling $\lambda_1$ from the SM value $\lambda$ is always positive and can be very large, even up to the order of 10, as expected in Table~\ref{hselfcpl} and Eq.~(\ref{eqn:diff}).
We can also read from the left and middle panels of Fig.~\ref{fig:random2} that a large deviation of the quartic coupling of SM-like Higgs need a relatively small $\rho_1$ and/or large $\alpha_1$.
As given in Eq.~(\ref{eqn:rhoT}), a large $y_N$ tends to decrease $\rho_T$, thus increasing the value of $v_c/T_c$. However, if $y_N$ is too large, say $y_N \gtrsim 1.5$,  a negative $\rho_T$ will be obtained which leads to a non-stable vacuum. Thus, the phase transition and GW in the LRSM favor a Yukawa coupling $y_N \sim {\cal O}(0.1)$ to ${\cal O}(1)$.

On the experimental side, the combined results of di-Higgs searches can be found e.g. in Refs.~\cite{Sirunyan:2018ayu,Aad:2019uzh}. Data from LHC 13 TeV with a luminosity of $36$ fb$^{-1}$ only set a weak constraint $\lambda_{hhh}/\lambda^{SM}_{hhh}\in(-5, 12)$. The LHC 14 TeV with an integrated luminosity of 3 ab$^{-1}$ can probe the trilinear coupling of SM Higgs within the range of $\lambda_{hhh}/\lambda^{\rm SM}_{hhh}\in(0.7,1.3)$~\cite{Barger:2013jfa}, while the future 100 TeV collider with a luminosity of 30 ab$^{-1}$ can help to improve the sensitivity up to  $\lambda_{hhh}/\lambda^{\rm SM}_{hhh}\in(0.9,1.1)$~\cite{Kilian:2017nio}. However, this is not precise enough to see the deviation of trilinear coupling in the LRSM, which is of order $10^{-3}$ or smaller.

Although the quartic coupling measurements can not be greatly improved at hadronic colliders~\cite{Kilian:2017nio, Chen:2015gva}, a future muon collider with the center-of-mass energy of 14 TeV and a luminosity of 33 ab$^{-1}$ can probe a deviation of the quartic Higgs self-coupling at the level of $50\%$~\cite{Chiesa:2020awd}. This can probe a sizable region of parameter space in Fig.~\ref{fig:random2}.


\subsection{Searches of $H_3^0$ and RHNs in the LRSM}
\label{sec:H3}

\begin{figure}[!t]
	\centering	
	\subfigure{\includegraphics[width=0.55\linewidth]{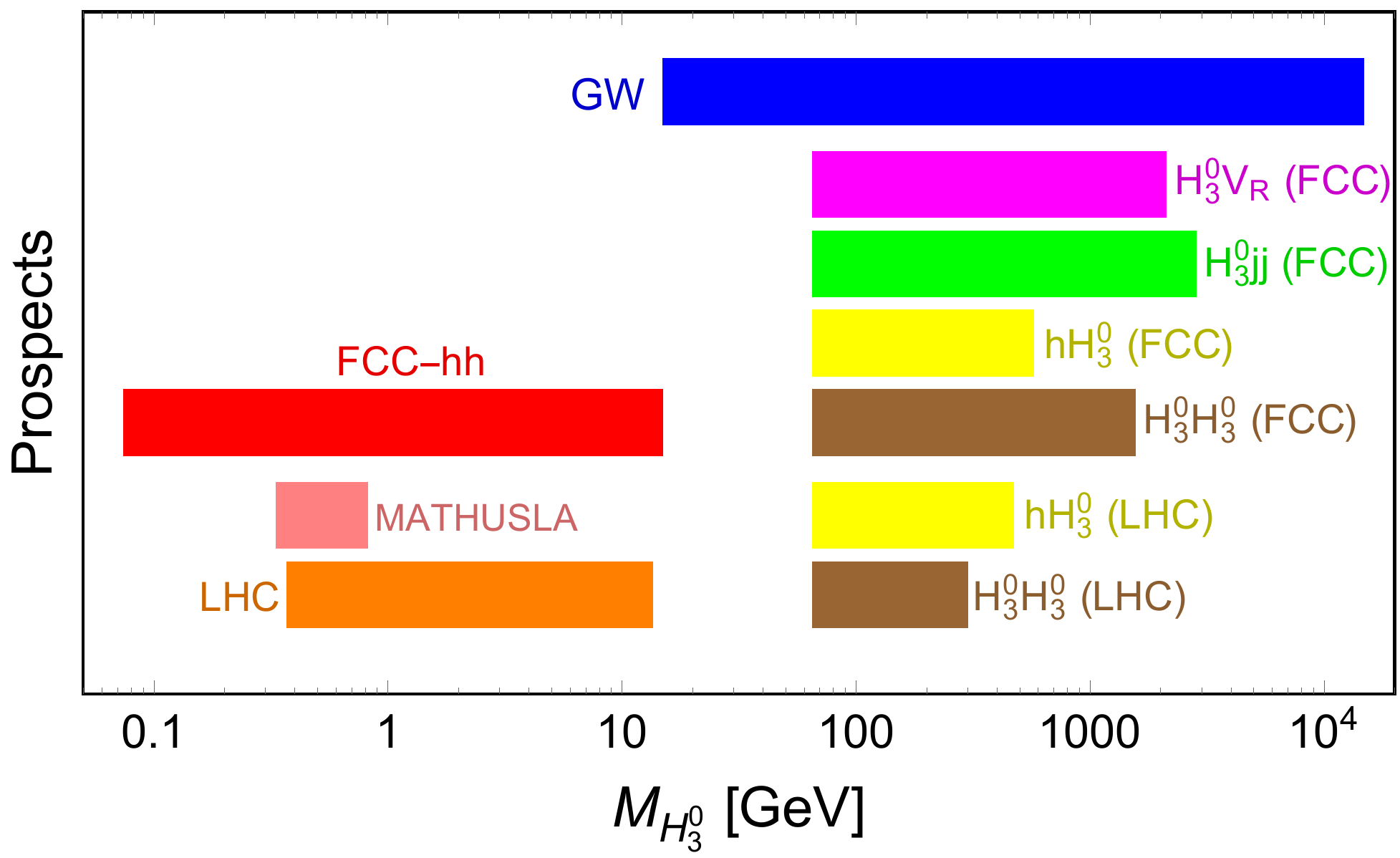}}
	\caption{Complementarity of $H_3^0$ at the colliders and GWs: the orange, pink and red bands are the prospects of a light $H_3^0$ at FCC-hh (100 TeV and 30 ab$^{-1}$), MATHUSLA and LHC (14 TeV and 3 ab$^{-1}$), the brown, yellow, green and magenta bands are the prospects of direct searches of $H_3^0$ at the FCC-hh (and LHC) in the channels $H_3^0 H_3^0$, $h H_3^0$, $H_3^0jj$ and $H_3^0V_R$. The blue band is the GW prospect of $H_3^0$ mass in the right panel of Fig.~\ref{fig:massofsamples}. }
	\label{fig:complementarity}
\end{figure}

As implied by the BPs in Figs.~\ref{fig:massofsamples} and \ref{fig:GWcurves}, the GW signals favor a relatively light $H_3^0$ in the LRSM, and this can be correlated to the direct searches of a (light) $H_3^0$ at the high-energy frontier. At the high-energy colliders, the scalar $H_3^0$ can be produced in two portals~\cite{Dev:2016dja}:
\begin{itemize}
	\item The scalar portal, i.e. the production of $H_3^0$ through its coupling to the SM Higgs $h$. This includes the channels $pp \to h^\ast \to h H_3^0$ and $pp \to h^{(\ast)} \to H_3^0 H_3^0$. The production amplitudes in both the two channels are proportional to the quartic coupling $\alpha_1$. As the trilinear couplings $\lambda_{hH_3^0 H_3^0}$ and $\lambda_{hh H_3^0}$ are  respectively proportional to the VEVs $v_{\rm EW}$ and $v_R$, even if $\alpha_1$ is small say $\alpha_1 \sim 10^{-2}$, the production cross sections are still sizable. Assuming $\alpha_1= 0.01$ and $v_R = 10$ TeV, the prospects of $H_3^0$ at the LHC 14 TeV with an integrated luminosity of 3 ab$^{-1}$ and the future 100 TeV collider with a luminosity of 30 ab$^{-1}$ are shown as the yellow and brown bands in Fig.~\ref{fig:complementarity}~\cite{Dev:2016dja}.
	\item The gauge portal, i.e. the production of $H_3$ through its couplings to the heavy $W_R$ and $Z_R$ gauge bosons, in the  Higgsstrahlung process $pp \to V_R^\ast \to H_3^0 V_R$ (with $V_R = W_R,\; Z_R$) and the vector boson fusion (VBF) process $pp \to H_3^0 jj$. In light of the current direct LHC constraints on $W_R$ and $Z_R$ (see Section~\ref{sec:experimental}), the prospects of $H_3^0$ at the LHC in these channels are very limited, which however can be largely improved at future 100 TeV colliders. The FCC-hh prospects in the $H_3^0 jj$ and $H_3^0 V_R$ channels are shown respectively as the green and magenta bands in Fig.~\ref{fig:complementarity}.
\end{itemize}
In obtaining both the scalar and gauge portal prospects, we have set a lower bound on the $H_3^0$ mass, i.e.  $M_{H_3^0} > m_h/2\simeq 62.5$ GeV, such that the exotic decay of the SM Higgs $h \to H_3^0 H_3^0$ is kinematically forbidden~\cite{Curtin:2013fra}.

The scalar $H_3^0$ mixes with the SM Higgs $h$ and the heavy bidoublet scalar $H_1^0$, which induces the tree-level FCNC couplings of $H_3^0$ to the SM quarks. Therefore for sufficiently light $H_3^0$, it can be produced from flavor-changing meson decays, such as $K \to \pi H_3^0$~\cite{Dev:2016vle,Dev:2017dui}. The high-precision SM meson data have set very severe constraints on the mixing angles of $H_3^0$ with $h$ and $H_1^0$. Therefore in a large region of parameter space the light $H_3^0$ decays predominately into two photons $H_3^0 \to \gamma\gamma$ through the $W_R$ and heavy charged scalar loops in the LRSM. Suppressed  by the heavy particle masses in the loops, the scalar $H_3^0$ tends to be long-lived, and can thus be searched in the multi-purpose detectors at the high-energy colliders as well as the dedicated LLP experiments therein. The prospects of long-lived $H_3$ at the LHC 14 TeV, FCC-hh, and MATHUSLA~\cite{Curtin:2018mvb} are presented Fig.~\ref{fig:complementarity} respectively as the orange, red and pink bands~\cite{Dev:2016vle,Dev:2017dui}.

The GW prospect of $M_{H_3}$ in Fig.~\ref{fig:massofsamples} is indicated by the blue band in Fig.~\ref{fig:complementarity}. As clearly seen in Fig.~\ref{fig:complementarity}, the direct searches of $H_3^0$ at the LHC and future 100 TeV colliders can probe a mass range of roughly 100 GeV up to 3 TeV, while the searches of a long-lived $H_3^0$ at the high-energy colliders can cover the mass range of 10 GeV down to 100 MeV. As a new avenue to probe the phase transition in the LRSM, GWs are sensitive to a wide mass range of $H_3^0$, from the 10 GeV scale up to 10 TeV, which is largely complementary to the searches of (light) $H_3^0$ at the high-energy colliders.

Note that one of the important decay modes of $H_3^0$ is the RHN channel, i.e. $H_3^0 \to NN$, which will induce the strikingly clean signal of same-sign dilepton plus jets~\cite{Dev:2016dja,Maiezza:2015lza, Nemevsek:2016enw}. The heavy RHNs can also be produced through their gauge couplings to the $W_R$ and $Z_R$ bosons, e.g. the smoking-gun Keung-Senjanovi\'{c} signal $pp \to W_R \to N \ell^\pm \to \ell^\pm \ell^\pm jj$ at the high-energy $pp$ colliders~\cite{Keung:1983uu}. If the RHNs are very light, say below 100 GeV scale, the decay widths of RHNs will be highly suppressed by $W_R$ mass, which makes the RHNs long-lived~\cite{Helo:2013esa, Cottin:2018kmq}. The light long-lived RHNs can be searched directly at the high-energy colliders via displaced vertex, or even from meson decays~\cite{Helo:2010cw, Cvetic:2010rw, Drewes:2015iva, Bondarenko:2018ptm}. The prospects of RHNs at the high-energy colliders and in meson decays depend largely on the heavy scalar or gauge boson masses (see also~\cite{Mitra:2016kov, Ruiz:2017nip}).  However, it is worth pointing out that, as seen in Fig.~\ref{fig:massofsamples}, GWs are sensitive to the RHN masses in the range of 200 GeV up to 40 TeV, which is largely complementary to the direct searches of (light) RHNs at the high-energy frontier.

\section{Discussions and Conclusion}
\label{sec:conclusion}

Before the conclusion we would like to comment on some open questions in the phase transition and GW production in the LRSM:
\begin{itemize}
	\item In the calculations we have assumed that at the epoch of phase transition the bubbles expanding in the plasma can reach a relativistic terminal velocity, i.e. the non-runaway scenarios, where the velocity of bubble wall is taken to be $v_w \simeq 1 $ in our analysis, which corresponds to the denotation case \cite{Espinosa:2010hh}. A recent numerical analysis~\cite{Cutting:2019zws} has revealed that the SW contribution might be suppressed by a factor of $10^{-3}$ in the deflagration case when $\alpha > 0.1$ where the reheated droplet can suppress the formation of GW signals. While there is no such a huge suppression for the denotation case with $\alpha<0.1$, our results could still be valid, although the GW signals might be suppressed by a factor two or three.
	The bubble wave velocity, in principle, can be computed from the parameters of a given model, as demonstrated in \cite{Moore:1995ua, Moore:1995si, Bodeker:2009qy}. Furthermore, according to the recent calculations in Ref.~\cite{Guo:2020grp}, it is found that the finite lifetime of SWs can lead to a suppression factor $\Upsilon$, which can be parameterized in the following form~\cite{Fornal:2020esl} 
	\begin{eqnarray}
		\label{eqn:upsilon}
		\Upsilon = 1-\left[1+\frac{8\pi^{1/3}}{\sqrt{3}}v_{w}\frac{H_*}{\beta}\left(\frac{\alpha\kappa_v}{1+\alpha}\right)^{-1/2}\right]^{-1/2}\,.
	\end{eqnarray}
	We have calculated the $\Upsilon$ factors for the five BPs in Table~\ref{tab:BPs}, and listed it in the last row of this table. It is observed that the GW signals in these BPs might be suppressed by up to a factor of 6 to 10. 
	It might be interesting to explore how the model parameters of LRSM can affect the bubble wall velocity and the effects of the suppression factor $\Upsilon$, which will be a topic for our future study.
	
	\item It is remarkable that for the scalar $H_3^0$, which is mainly the  CP-even neutral component of the right-handed triplet $\Delta_R$, both the theoretical and experimental constraints on it are very weak. As a result, its mass could span a wide range, say from below GeV-scale up to tens of TeV. In the case that all other new particles in the LRSM are heavier than 5 TeV but with a relatively light $H^0_3$ below the TeV-scale (for instance the BPs BP1 and BP2 in Table~\ref{tab:BPs}), at the scale below 1 TeV, the scalar potential of LRSM given in Eq. (\ref{pot}) can be reduced to the effective model with the SM extended by a real singlet $S$, where the scalar potential has the following form:
	\begin{eqnarray}
		V(H, S) & \ = \ & -\mu^2 (H^\dagger H) + \frac{1}{2} m_S^2 S^2
		+ \frac{1}{4} \lambda (H^\dagger H)^2  +  \lambda_{3S} S^3 + \lambda_{4S} S^4 \nonumber \\
		&& + \lambda_{3X} S (H^\dagger H) + \lambda_{4X} S^2 (H^\dagger H) \,.
		\label{xsm}
	\end{eqnarray}
	The trilinear and quartic couplings in Eq.~(\ref{xsm}) can be written as functions of the right-handed VEV $v_R$ and the quartic couplings in the LRSM, which are collected in Table~\ref{xsmcpl}.
	Obviously, when $\alpha_1$ is switched off, $H^0_3$ will not affect the EW phase transition directly, and the EW phase transition should be of second-order as in the SM. When $\alpha_1$ is switched on, it might be interesting to examine whether the light $H_3^0$ can affect the phase transitions at both the $v_R$ scale and the EW  scale. When it is possible, a multi-step strong FOPT could be expected \cite{Angelescu:2018dkk}.
\end{itemize}

\begin{table}[!]
	\begin{center}
		\caption{Trilinear and quartic couplings given in Eq. (\ref{xsm}) for a SM+singlet model derived from the LRSM. \label{xsmcpl}}
		\begin{tabular}{|c|c|}
			\hline
			trilinear couplings &  expressions \\
			\hline
			$\lambda_{3S}$ & $ \sqrt{2} \rho_1 v_R$ \\
			\hline
			$\lambda_{3X}$ & $\frac{1}{\sqrt{2}} \alpha_1 v_R$ \\
			\hline
			\hline
			quartic couplings &  expressions \\
			\hline
			$\lambda$ & $\lambda_1$\\
			\hline
			$\lambda_{4S}$ &  $\frac{1}{4} \rho_1$ \\
			\hline
			$\lambda_{4X}$ & $\frac{1}{2} \alpha_1$ \\
			\hline
		\end{tabular}
	\end{center}
\end{table}


To summarize, in this paper we have studied the prospects of GW signals from phase transition in the minimal LRSM with a bidoublet $\Phi$, a left-handed triplet $\Delta_L$ and a right-handed triplet $\Delta_R$, which is a well-motivated framework to restore parity and accommodate the seesaw mechanisms for tiny neutrino masses at the TeV-scale. We have considered the theoretical limits on the LRSM from perturbativity, unitarity, vacuum stability and correct vacuum criteria, as well as the experimental constraints on the heavy gauge bosons and the BSM scalars. The experimental limits are collected in Table~\ref{table:bounds} and Fig.~\ref{fig:spectra}.

With these theoretical and experimental constraints taken into account, we have analyzed the parameter space of strong FOPT and the resultant GWs in the LRSM. As demonstrated in Figs.~\ref{figvT}, \ref{fig:random1} and \ref{fig:random2}, the strong FOPT at the $v_R$ scale favors relatively small quartic and Yukawa couplings, which corresponds to relatively light BSM scalars and RHNs. The GWs for some BPs in the LRSM in Fig.~\ref{GWpeak} reveal that the phase transition in the LRSM can generate the GW signals of $10^{-17}$ to $10^{-12}$, with a frequency ranging from 0.1 to 10 Hz, which can be probed by the experiments BBO and DECIGO, or even by ALIA and MAGIS. Setting $v_R = 10$ TeV, as seen in Fig.~\ref{fig:massofsamples}, the GWs are sensitive to the following mass ranges:
\begin{itemize}
	\item The heavy bidoublet scalars $H_1^0$, $A_1^0$, $H_1^\pm$, the scalars $H_2^0$, $A_2^0$, $H_2^\pm$ and $H_1^{\pm\pm}$ from the left-handed triplet $\Delta_L$, and the doubly-charged scalar $H_2^{\pm\pm}$ from the right-handed triplet $\Delta_R$, with masses up to tens of TeV, with the lower bounds of their masses roughly set by the experimental limits in Fig.~\ref{fig:spectra}.
	\item The scalar $H_3^0$ with mass in the range of roughly from 20 GeV up to 10 TeV. As presented in Fig.~\ref{fig:complementarity}, the GW prospects of $H_3^0$ are largely complementary to the direct searches of heavy $H_3^0$ at the LHC  and future 100 TeV colliders, and the searches of light $H_3^0$ from displaced vertex signals at the LHC, future higher energy colliders, and the LLP experiments such as MATHUSLA.
	\item The RHNs with masses from roughly 300 GeV up to 40 TeV. The GW sensitivity of $M_N$ is also largely complementary to the direct searches of prompt signals and displaced vertices from RHNs at the high-energy colliders, as well as the production of RHNs from meson decays.
\end{itemize}
The GW spectra in Fig.~\ref{fig:GWcurves} for the BPs in Table~\ref{tab:BPs} shows that the quartic coupling $\rho_1$ is crucially important for both the frequency and strength of the GW signals in the LRSM, while other couplings such as $\rho_2$, $\rho_3-2\rho_1$, $\alpha_3$ and $y_N$ are also important. In addition, the precision measurement of the quartic coupling of the SM Higgs at a future muon collider can probe a sizable region of the parameter space in LRSM, which can have strong FOPT and observable GW signals, as exemplified in Fig.~\ref{fig:random2}.


\section*{Acknowledgments}
This work is supported by the Natural Science Foundation of China under the grant no. 11575005. Y.Z. would like to thank P. S. Bhupal Dev and Yiyang Zhang for the helpful discussions at the early stage of this paper. The authors would also like to thank Dr. Yi-Dian Chen, Dr. Huaike Guo, Dr. Bartosz Fornal, Dr. White Graham Albert, and Dr. Zhi-Wei Wang for some useful information.

\appendix
\section{Mass matrices and thermal self-energies}
\label{appendix:masses}

\begin{table}[!t]
	\begin{center}
		\caption{Physical Higgs states and their masses when $v_L\ll\kappa_2\ll\kappa_1\ll v_R$~\cite{Zhang:2007da}. Here $\xi = \kappa_2/\kappa_1$ $\epsilon = v_{\rm EW}/v_R \simeq \kappa_1 / v_R$. $h$ is the SM Higgs field.}
		\label{tab:mass}		
		\begin{tabular}{ll}
			\hline\hline
			physical states & mass squared \\ \hline
			$h =\sqrt{2}\text{Re}[\phi_1^{0*}+\xi \phi_2^0] + \frac{\alpha_1 \epsilon}{\sqrt2 \rho_1} {\rm Re} [\Delta_R^0]$ &$\frac{1}{2}(4\lambda_1-\frac{\alpha_1^2}{\rho_1})v_{\rm EW}^2$  \\ \hline
			
			$H_1^0=\sqrt{2}\text{Re}[\phi_2^{0}-\xi \phi_1^{0*}]$ & \multirow{3}{*}{$\frac12 \alpha_3 v_R^2$} \\
			$A_1^0=\sqrt{2}\text{Im}[\phi_2^{0}-\xi \phi_1^{0*}]$ & \\
			${H_1^\pm} =\phi_2^{\pm}+\xi \phi_1^\pm + \frac{\epsilon}{\sqrt{2}}\Delta_R^\pm$& \\ \hline
			
			${H_2^0} = \sqrt{2}\text{Re}[\Delta_L^0]$ &
			\multirow{4}{*}{$\frac12 (\rho_3-2\rho_1) v_R^2$} 	\\
			$A_2^0=\sqrt{2}\text{Im}[\Delta_L^0]$ & \\
			${H_2^\pm} = \Delta_L^\pm$ & \\
			$H_1^{\pm\pm} = \Delta_L^{\pm\pm}$ & \\ \hline
			
			${H_3^0} = \sqrt{2}\text{Re}[\Delta_R^0]$&$2\rho_1 v_R^2$\\
			
			$H_2^{\pm\pm} = \Delta_R^{++}$ & $2\rho_2v_R^2$ \\	\hline\hline
		\end{tabular}
	\end{center}
\end{table}

In the LRSM with a bidoublet $\Phi$, a left-handed triplet $\Delta_L$ and a right-handed triplet $\Delta_R$, there are 20 degrees of freedom in the scalar sector. In this paper, for simplicity we assume there is no CP violation in the scalar sector, i.e. the CP phase $\delta=0$ in the potential (\ref{eqn:potential}) and the phases $\theta_\kappa = \theta_L = 0$ in the VEVs (\ref{eqn:vev}). In the limit of $v_L\ll\kappa_2\ll\kappa_1 \simeq v_{\rm EW} \ll v_R$, all the physical scalars and their masses are collected in Table~\ref{tab:mass}. The corresponding mass matrix elements can be found e.g. in Ref.~\cite{Deshpande:1990ip}. In the basis of $\sqrt{2}\{\text{Re}[\phi_1^0] ,\text{Re}[\phi_2^0],\text{Re}[\delta_L^0], \text{Re}[\delta_R^0]\}$, the thermal self-energy of the real neutral components are respectively:
\begin{eqnarray}
\left( \Pi_{H^0}\right)_{11} & \ = \ & \left( \Pi_{H^0}\right)_{22} \ = \ \frac{T^2}{24} \left( \frac92g_L^2 + \frac92 g_R^2 + 20\lambda_1+8\lambda_3+12\alpha_1+6\alpha_3+6y_t^2+6y_b^2 \right) \,, \nonumber \\ && \\
\left( \Pi_{H^0}\right)_{33} & \ = \ & \frac{T^2}{24} \left( 12g_L^2+6g_{BL}^2+16\rho_1+8\rho_2+6\rho_3+8\alpha_1+4\alpha_3+12y_N^2 \right) \,, \\
\left( \Pi_{H^0}\right)_{44} & \ = \ & \frac{T^2}{24} \left( 12g_R^2+6g_{BL}^2+16\rho_1+8\rho_2+6\rho_3+8\alpha_1+4\alpha_3+12y_N^2 \right) \,, \\
\left( \Pi_{H^0}\right)_{12}& \ = \ & T^2(\alpha_2+\lambda_4+y_t y_b) \,, \\
\left( \Pi_{H^0}\right)_{13}& \ = \ & \left( \Pi_{H^0}\right)_{14} \ = \ \left( \Pi_{H^0}\right)_{23} \ = \ \left( \Pi_{H^0}\right)_{24} \ = \ \left( \Pi_{H^0}\right)_{34} \ = \ 0 \,.
\end{eqnarray}
All the rest elements are related to the ones above via $(\Pi_{H^0})_{ij} = (\Pi_{H^0})_{ji}$. The thermal self-energy for the imaginary components of the neutral scalars is very similar to that for the real components.  In the  basis of $\sqrt{2}\{\text{Im}[\phi_1^0] ,\text{Im}[\phi_2^0],\text{Im}[\delta_L^0], \text{Im}[\delta_R^0]\}$, the elements are respectively:
\begin{eqnarray}
\left( \Pi_{A^0}\right)_{ij} \ = \
\begin{cases}
	+\left( \Pi_{H^0}\right)_{ij}, & \text{for } (i,j)\ne (1,2) \,, \\
	-\left( \Pi_{H^0}\right)_{ij}, & \text{for } (i,j)= (1,2) \,.
\end{cases}
\end{eqnarray}
For the singly charged fields, in the basis of  $\{\phi_1^\pm ,\phi_2^\pm,\delta_L^\pm,\delta_R^\pm \}$, the thermal self-energy is the same as that for the real neutral components, i.e. $\Pi_{H^\pm}=\Pi_{H^0}$. For the doubly-charged scalars, in the basis of $\{\Delta_L^{\pm\pm},\Delta_R^{\pm\pm}\}$, the corresponding self-energy is  given by
\begin{eqnarray}
\left( \Pi_{H^{\pm\pm}}\right)_{11}= \left( \Pi_{H^{0}}\right)_{33} \,, \quad
\left( \Pi_{H^{\pm\pm}}\right)_{22}=\left( \Pi_{H^{0}}\right)_{44} \,, \quad
\left( \Pi_{H^{\pm\pm}}\right)_{12}=0 \,.
\end{eqnarray}

For the neutral gauge bosons, in the basis of $\{W_L^3,W_R^3,B\}$, the self-energy matrix reads
\begin{eqnarray}
	\Pi_{W_3B}=\frac{T^2}{6}\text{diag}\{ 9g_L^2,9g_R^2,17g_{BL}^2 \} \,,
\end{eqnarray}
while for the singly-charged gauge bosons, in the basis of $\{W_L^\pm,W_R^\pm\}$, the self-energy matrix is
\begin{eqnarray}
	\Pi_{W^\pm}=\frac{3T^2}{2}\text{diag} \{ g_L^2,g_R^2 \} \,.
\end{eqnarray}





\section{Conditions for vacuum stability and correct vacuum}
\label{appendix:vacuum}

The sufficient but not necessary conditions for vacuum stability and correct vacuum in the LRSM are worked out in \cite{Chauhan:2019fji} and listed below (simple analytic formula can only be obtained in the condition $\alpha_2=0$):
\begin{eqnarray}
\label{eqn:correctvacuum}
&&\rho_1 > 0\,,\,\, \rho_2 > 0\,,\,\, \rho_3 > 2 \rho_1\,,\,\,|\rho_4| < \frac{\rho_3 - 2 \rho_1}{2} + \rho_2 \,, \nonumber \\
&&\alpha_1 +  2 \sqrt{\lambda_1 \rho_1} > 0\,,\,\,\alpha_1 +  \alpha_3 + 2 \sqrt{\lambda_1 \rho_1} > 0\,,\,\, \nonumber \\
&&\alpha_1 + \frac{\alpha_3}{2} \left ( 1 \pm \sqrt{1 - \frac{\lambda_4^2}{(2 \lambda_2 + \lambda_3)^2}} \right ) + 2 \sqrt{\left(\lambda_1 - \frac{\lambda_4^2}{2 \lambda_2 + \lambda_3}\right) \rho_1} > 0\,, \nonumber \\
&&\alpha_1 + \frac{\alpha_3}{2} + 2 \sqrt{\left(\lambda_1 + \lambda_3 - 2 \lambda_2 - \frac{\lambda_4^2}{4 \lambda_2}\right) \rho_1} \, > 0\,, \nonumber \\
&&\alpha_1 + \frac{\alpha_3}{2} + 2 \sqrt{(\lambda_1 + \lambda_3 + 2 ( \lambda_2 - |\lambda_4|)  \rho_1} \, > 0\,, \nonumber \\
&&2\sqrt{\text{min}[f_{\rm SSB}]\rho_1}-\left|\left|\alpha_1+\frac{\alpha_3}{2}\left(1-\text{sign}(\alpha_3)\sqrt{1-\eta^2}\right)\right|\right|>0\,, \nonumber \\
&&2\text{min}[\rm f_{SSB}]\mu_3^2-\left[\alpha_1+\frac{\alpha_3}{2}\left(1-\text{sign}(\alpha_3)\sqrt{1-\eta^2}\right)\right]\bar{\mu}_1^2>0\,, \nonumber \\
&&2\rho_1\bar{\mu}_1^2-\left[\alpha_1+\frac{\alpha_3}{2}\left(1-\text{sign}(\alpha_3)\sqrt{1-\eta^2}\right)\right]\mu_3^2>0 \,,
\end{eqnarray}
where $\bar{\mu}_1^2 \equiv \mu_1^2+2\sigma\mu_2^2$, with the definition  $\eta e^{i\omega} \equiv {{\rm Tr} [\tilde{\Phi} \Phi^\dagger]}/{{\rm Tr} [{\Phi}^\dagger \Phi]}$,
the parameter $\sigma$ is defined via $\sigma = \eta \cos \omega$, and
\begin{eqnarray}
	f_{\rm SSB} = \left\{
	\begin{array}{ll}
		\lambda_1 > 0\,, & \eta =\sigma=0 \\
		\lambda_1 - \frac{\lambda_4^2}{2 \lambda_2 + \lambda_3} > 0 \,, & \Leftarrow 0 < \eta =\frac{\left|\lambda_4\right|}{2 \lambda_2 + \lambda_3}<1, \sigma=\frac{-\lambda_4}{2 \lambda_2 + \lambda_3} \\
		\lambda_1 + \lambda_3 + 2 (\lambda_2 - |\lambda_4| ) > 0 \,, & \eta =1, \sigma=-\text{sign}(\lambda_4)\\
		\lambda_1 + \lambda_3 - 2 \lambda_2 - \frac{\lambda_4^2}{4 \lambda_2} > 0 \,, & \Leftarrow 4\left|\lambda_2\right|>\left|\lambda_4\right|, \eta =1, \sigma=-\frac{\lambda_4}{4\lambda_2}\,,
	\end{array} \right.
\end{eqnarray}
where the condition structure ``$p \Leftarrow q$'' means $p$ needs to be checked if and only if the condition $q$ is true.
In this paper, we have chosen $\lambda_{2,3,4}=0$, which corresponds to the case of $\eta =\sigma=0$.

\bibliographystyle{JHEP}
\bibliography{mybib}

\end{document}